%% file: qcd2.tex
\documentclass[floatfix,nofootinbib,aps,prd,preprintnumbers,showpacs,showkeys]{revtex4-1}

\usepackage{amssymb}
\usepackage{amsmath}
\usepackage{color}
\usepackage{graphicx}% Include figure files
\usepackage{dcolumn}% Align table columns on decimal point
\usepackage{bm}% bold math
\usepackage{epsfig}

\def\figdir{}

\newcommand\figWidthHalf{.48\textwidth}

\newcommand\Eq[1]{Eq.~\ref{eq:#1}}
\newcommand\Fig[1]{Fig.~\ref{fig:#1}}
\newcommand\Sec[1]{Sec.~\ref{sec:#1}}

\newcommand\Tab[1]{Table~\ref{tab:#1}}
\newcommand\bfx{\mathbf x}
\newcommand\bfy{\mathbf y}
\newcommand\bfz{\mathbf z}

\newcommand\calO{\mathcal O}

\newcommand\calP{\mathcal P}

\newcommand\calM{\mathcal M}
\newcommand\Tr{\textrm{Tr}}
\newcommand\Transpose{\intercal}

\begin{document}

\preprint{MIT-CTP/4808}

\title{Multiscale Monte Carlo equilibration: Two-color QCD with two fermion flavors}

\author{William Detmold}
\email{wdeltmold@mit.edu}
\affiliation{Center for Theoretical Physics, Massachusetts Institute of Technology, Cambridge, Massachusetts 02139, USA}

\author{Michael G. Endres}
\email{endres@mit.edu}
\affiliation{Center for Theoretical Physics, Massachusetts Institute of Technology, Cambridge, Massachusetts  02139, USA}


\date{\today}

\begin{abstract}
We demonstrate the applicability of a recently proposed multiscale thermalization algorithm to two-color quantum chromodynamics (QCD) with two mass-degenerate fermion flavors.
The algorithm involves refining an ensemble of gauge configurations that had been generated using a renormalization group (RG) matched coarse action, thereby producing a fine ensemble that is close to the thermalized distribution of a target fine action; the refined ensemble is subsequently rethermalized using conventional algorithms.
Although the generalization of this algorithm from pure Yang-Mills theory to QCD with dynamical fermions is straightforward, we find that in the latter case, the method is susceptible to numerical instabilities during the initial stages of rethermalization when using the hybrid Monte Carlo algorithm.
We find that these instabilities arise from large fermion forces in the evolution, which are attributed to an accumulation of spurious near-zero modes of the Dirac operator.
We propose a simple strategy for curing this problem, and demonstrate that rapid thermalization--as probed by a variety of gluonic and fermionic operators--is possible with the use of this solution.
In addition, we study the sensitivity of rethermalization rates to the RG matching of the coarse and fine actions, and identify effective matching conditions based on a variety of measured scales.

\end{abstract}

\maketitle

\section{Introduction}
\label{sec:introduction}

Multilevel methods have played an increasingly important role in various aspects of lattice QCD simulations, ranging from Dirac operator inversion~\cite{Babich:2010qb,Babich:2009pc,Frommer:2013fsa,Brannick:2014vda} to evaluation of correlation functions and other observables~\cite{Luscher:2001up,Ce:2016idq,Vera:2016xpp}.
Although multilevel methods have been applied successfully to Monte Carlo updating in simple systems~\cite{Goodman:1986pv,Edwards:1990hu,Edwards:1991eg,Janke:1993et,Grabenstein:1993nh,Grabenstein:1994ze}, generalization to gauge evolution in QCD remains an open challenge.
Such methods offer the prospect of dramatically speeding up lattice QCD calculations as they are exponentially more efficient in exploring the space of configurations and avoid the critical slowing down that current algorithms typically exhibit as the continuum limit is approached.
Recently, a multiscale method was proposed, which combines standard gauge evolution techniques with the multigrid notions of prolongation and refinement to achieve rapid thermalization of configurations for pure Yang-Mills theory~\cite{Endres:2015yca}.
With such methods at hand, it was demonstrated that fully decorrelated streams of ensembles with well-sampled topology could be efficiently generated, even in the regime of fine lattice spacing, where topological freezing can be problematic~\cite{PhysRevD.73.094507,Schaefer:2010hu}.

The multilevel thermalization method introduced in~\cite{Endres:2015yca} involves several steps, which we briefly review.
First, a decorrelated coarse ensemble is generated using a renormalization group (RG) matched coarse action.
Subsequently, the coarse configurations are prolongated, or refined, to produce a fine ensemble.
Finally, the configurations in the prolongated ensemble are evolved (as parallel streams) according to the fine action until the ensemble equilibrates.
Several key ingredients make this algorithm more efficient than conventional algorithms, which rely upon a single Markov chain to generate decorrelated configurations.
At sufficiently fine lattice spacing, the prolongator, appropriately defined, preserves the topological charge density (and therefore topological charge) on a per-configuration basis.
Thus, the prolongated ensemble inherits the topological charge distribution of the coarse ensemble, which, if produced using a properly matched coarse action, will be correctly distributed according to the probability measure defined by the fine action up to (coarse) discretization artifacts.
In addition, provided the coarse action is properly tuned, the long-distance character of the prolongated ensemble will reflect that of an ensemble generated using the fine action, and only the short-distance properties will require correction.
The latter motivates the final stage of fine evolution, which allows the prolongated ensemble to return to equilibrium.
Given that it is the short-distance properties of the prolongated ensemble that require correction, it is reasonable to expect that the ensemble would return to equilibrium rapidly, even when using a local updating algorithm.

The utility of multilevel thermalization was established for pure Yang-Mills theory for the gauge group $SU(3)$~\cite{Endres:2015yca}.
At first glance, the introduction of fermions poses no apparent additional complication to this strategy, although the presence of fermion mass scales make the RG matching of the coarse action more involved.
As will be demonstrated, however, the presence of spurious near-zero modes of the fine Dirac operator make the initial evolution of prolongated ensembles unstable in the presence of fermions, and additional measures are required to cure the problem.
We investigate these issues in detail for two-color QCD with two flavors of heavy fermions, and demonstrate that with modifications to the multilevel thermalization strategy, rapid thermalization can be achieved even for ensembles generated with dynamical fermions.
As a byproduct of this study, we observe that the shortest rethermalization times are indeed achieved when the coarse and fine actions are appropriately RG matched.
This empirical observation bridges the disconnect between the algorithm dependence of rethermalization and the underlying physical intuition that motivates a multiscale strategy.

Having constructed an effective multilevel strategy, we explore possible ways in which the RG matching of the coarse ensemble can be performed and determine the efficacy of the matching through investigations of the rethermalization cost of a range of short- and long-distance quantities.
These include observables constructed from gluonic degrees of freedom (e.g., functions of Wilson loops) as well as from fermionic degrees of freedom (e.g., pion and rho meson correlation functions).
It is apparent from the results that tuning is nontrivial to perform perfectly, with different observables leading to different predictions for the RG matched coarse-action parameters, but in the case at hand, a reasonable best guess could be determined, with {\it post facto} rethermalization studies showing the effectiveness of the choice.
The challenges encountered with RG matching are expected to diminish with lattice spacing, however, as eventually the matching will become controlled by perturbative QCD.

\section{Action, Observables, Ensembles, and Refinement}
\label{sec:ensembles}

We consider two-color QCD with two fermion flavors on an isotropic four-dimensional Euclidean space-time lattice, with lattice spacing $a$.
The action is given by $S=S_g + S_f$, where $S_g$ and $S_f$ represent the gauge and fermion contributions to the action, respectively.
For simplicity, we consider the Wilson (plaquette) gauge action~\cite{PhysRevD.10.2445}, given by
\begin{eqnarray}
S_g = -\frac{\beta}{2} \sum_x\sum_{\mu<\nu} W_{\mu\nu}(x) \ ,
\end{eqnarray}
where
\begin{eqnarray}
W_{\mu\nu}(x) = \frac{1}{2} \Re \Tr \left[ U_\mu(x) U_\nu(x+a e_\mu) U_\mu(x+a e_\nu)^\dagger U_\nu(x)^\dagger \right]\ ,
\end{eqnarray}
the link variables $U_\mu(x) \in SU(2)$ for all orientations $\mu = 0,1,2,3$ and positions $x$, $e_\mu$ is a basis vector oriented in the $\mu$-direction, $\beta = 4/g_0^2$, and $g_0$ is the bare gauge coupling.
The fermion action is given by
\begin{eqnarray}
S_f = \sum_{f=u,d} \bar\psi_f D_w(m_0) \psi_f\ ,
\end{eqnarray}
where $D_w(m_0)$ is the Wilson-Dirac operator evaluated at a flavor-independent bare quark mass $m_0$ (i.e., we work in the isospin symmetric limit).
The Wilson-Dirac operator is given explicitly by
\begin{eqnarray}
D_w(m_0) = \left(\frac{4}{a}+m_0\right) - \frac{1}{a}\sum_{\mu=0}^3 \left( P_\mu^- \Omega_\mu^+ + P_\mu^+ \Omega_\mu^- \right) \ ,
\end{eqnarray}
where
\begin{eqnarray}
P_\mu^{\pm} = \frac{1}{2} (1 \pm \gamma_\mu) \qquad \langle x | \Omega_\mu^+ | y \rangle = \delta_{x+\mu,y}U(x,\mu) 
\end{eqnarray}
and  $\Omega_\mu^-= (\Omega_\mu^+)^\dagger$.

Numerical studies were performed on lattices of space-time volume $V=L^3\times T$, taking $T=3L$ in order to minimize thermal effects and enable reliable spectroscopic measurements.
Periodic boundary conditions were imposed in all directions for the gauge field, whereas (anti)periodic boundary conditions in the (time) space directions were imposed for the fermions.
A summary of the coarse and fine simulation parameters considered can be found in \Tab{ensembles_generation}.
All simulations were performed in double precision using the hybrid Monte Carlo (HMC) updating algorithm~\cite{DUANE1987216}, with trajectories of unit length produced using a standard leapfrog algorithm.
A residual stopping condition of $10^{-10}$ was used throughout for the Dirac operator inversions.
For simplicity, we choose to use the same functional form for coarse and fine actions but note that an improved RG matching would be possible with a more complicated coarse action.

\begin{table}
\caption{%
\label{tab:ensembles_generation}%
Coarse and fine ensemble generation parameters: $N_\textrm{traj}$ represents the total number of HMC trajectories,  $N_\textrm{ckpt}$ represents the frequency with which configurations are stored for subsequent measurement, $N_\textrm{eq}$ represents the number of trajectories generated before measurements take place, $1/\delta\tau$ is the total number of steps per trajectory, and $P_\textrm{acc}$ is the acceptance probability associated with the accept/reject step at the end of each trajectory.
All ensembles were generated using unit length trajectories.
The total number of thermalized configurations stored is given by $N_\textrm{conf} = (N_\textrm{traj} - N_\textrm{eq})/N_\textrm{ckpt}$.
}
\begin{ruledtabular}
\begin{tabular}{ccccccccc}
Label & Lattice           & $\beta$ & $a m_0$ & $N_\textrm{traj}$ & $N_\textrm{ckpt}$  & $N_\textrm{eq}$ & $\delta\tau$ & $P_\textrm{acc}$ (\%) \\
\hline
$C_1$    & $12^3\times 36$   & 1.750 & -1.0000 & 9620  & 10 & 500  & 1/35 & 83  \\
$C_2$    & $12^3\times 36$   & 1.750 & -1.1760 & 5830  & 10 & 500  & 1/50 & 81  \\
$C_3$    & $12^3\times 36$   & 1.800 & -1.0000 & 11030 & 10 & 500  & 1/35 & 84  \\
$C_4$    & $12^3\times 36$   & 1.800 & -1.0500 & 11500 & 10 & 500  & 1/35 & 82  \\
$C_5$    & $12^3\times 36$   & 1.800 & -1.0890 & 5360  & 10 & 500  & 1/35 & 80  \\
$C_6$    & $12^3\times 36$   & 1.900 & -0.9000 & 10210 & 10 & 500  & 1/35 & 83  \\
$C_7$    & $12^3\times 36$   & 1.900 & -0.9500 & 6420  & 10 & 500  & 1/40 & 86  \\
$C_8$    & $12^3\times 36$   & 1.900 & -1.0000 & 8840  & 10 & 500  & 1/45 & 88  \\
$C_9$    & $12^3\times 36$   & 1.920 & -0.9700 & 5210  & 10 & 500  & 1/45 & 88  \\
$C_{10}$ & $12^3\times 36$   & 1.935 & -0.9350 & 5640  & 10 & 500  & 1/45 & 90  \\
$C_{11}$ & $12^3\times 36$   & 1.950 & -1.0000 & 5990  & 10 & 500  & 1/55 & 79  \\
$C_{12}$ & $12^3\times 36$   & 2.000 & -0.8500 & 6420  & 10 & 500  & 1/50 & 91  \\
$F$      & $24^3\times 72$   & 2.200 & -0.7200 & 7180  & 10 & 500  & 1/75 & 82  \\
\end{tabular}
\end{ruledtabular}
\end{table}

For the purpose of RG matching and/or subsequent thermalization studies, we study a variety of short- and long-distance quantities.
Short-distance observables under consideration include the space-time averaged plaquette, given by
\begin{eqnarray}
\langle \bar W \rangle = \frac{a^4}{6 V} \sum_x\sum_{\mu<\nu} \langle W_{\mu\nu}(x) \rangle
\end{eqnarray}
and bare (unrenormalized) chiral condensate,~\footnote{Note that the bare chiral condensate is dominated by UV divergent contributions, is not considered a long-distance observable in the context of this work, and therefore is not used for the purpose of RG matching. The Banks-Casher relation~\cite{Banks:1979yr} provides an alternative definition of the chiral condensate, expressed in terms of the lowest modes of the Dirac operator, and is therefore insensitive to such divergences; we will return to Dirac spectra and its relevance for RG matching in a later section.} given by
\begin{eqnarray}
\langle \bar\psi\psi \rangle^{\rm (bare)} = \frac{1}{8V} \langle \Tr D^{-1}_w(m_0) \rangle\ .
\end{eqnarray}
The space-time trace in the latter observable is evaluated stochastically using a different Gaussian random source for each spinor and color component~\cite{Edwards:1998wx}.
Extracted values and uncertainties for these quantities are provided in \Tab{ensembles_measurements} for each 
coarse and fine ensemble that we consider.

To probe long-distance properties of the theory, we consider vanishing three-momentum projected correlation functions constructed from isovector meson interpolating operators, given by 
\begin{eqnarray}
C_{\Gamma^\prime\Gamma}^{s^\prime s}(\tau) = \sum_\Delta \sum_\bfx \langle \bar \calO_{\Gamma^\prime}^{s^\prime}(\tau +\Delta,\bfx)  \calO_\Gamma^s(\Delta,\bfy) \rangle\ ,
\end{eqnarray}
where $x=\{\tau,\bfx\}$,
\begin{eqnarray}
\calO^s_\Gamma(\tau, \bfx) = \bar\Psi^s_u(\tau,\bfx) \Gamma \Psi^s_d(\tau,\bfx) \ ,
\end{eqnarray}
and
\begin{eqnarray}
\Psi^s_f(\tau,\bfz) = \sum_{\bfy} \phi_s(\tau,\bfx,\bfy)  \psi_f(\tau,\bfy)\ .
\end{eqnarray}
Note that $\bar \calO_\Gamma(\bar\psi,\psi) = \calO_\Gamma(\psi,\bar\psi)$, $\Gamma = \{1,\gamma_\mu, \gamma_{\mu\nu}, \gamma_{\mu5}, \gamma_5 \}$, $\Gamma^\dagger = \Gamma$ and $\Tr(\Gamma\Gamma^\prime)=4\delta_{\Gamma\Gamma^\prime}$.~\footnote{Recall that in Euclidean space, $\gamma_\mu = \gamma_\mu^\dagger = \gamma_\mu^{-1}$, $\{\gamma_5, \gamma_\mu\} = 0$, $\gamma_5^\dagger = \gamma_5$, $\gamma_{\mu\nu} = -\frac{i}{2}[\gamma_\mu,\gamma_\nu]$, and $\gamma_{\mu5}=i\gamma_\mu \gamma_5$.}
Two types of gauge covariant quark source wave functions were considered, namely point sources ($s=P$), given by $\phi_P(\tau, \bfx,\bfy) = \delta_{\bfx,\bfy}$, and (approximate) Gaussian smeared sources ($s=S$), given by $\phi_S(\tau,\bfx,\bfy) = \langle \bfx|  (1 - \kappa_G \Delta_\tau )^{N_G} | \bfy \rangle$, where $\kappa_G = w_G^2/(4 {N_G})$ ($w_G=4.0$ and $N_G=80$) and $\Delta_\tau$ is the three-dimensional gauge-covariant Laplacian~\cite{Albanese:1987ds}.
The Laplacian involves gauge links associated with time slice $\tau$ that have been stout smeared (10 applications) using a smearing factor equal to 0.08~\cite{Morningstar:2003gk}.

In the regime $\tau \gg \delta^{-1}$, where $\delta$ is the mass gap, and $\tau\ll T \to \infty$, the correlation functions defined above behave as
\begin{eqnarray}
C_{\Gamma^\prime\Gamma}^{s^\prime s}(\tau) \sim Z^{s^\prime}_{\Gamma^\prime} Z^s_\Gamma \left[ e^{-m_{\Gamma^{\prime}\Gamma} \tau} \pm e^{-m_{\Gamma^{\prime}\Gamma} (T-\tau)}\right]\ ,
\end{eqnarray}
up to exponentially suppressed excited state contamination.
A variety of scales may be extracted from these correlators at late times.
The pseudoscalar ($\Gamma=\gamma_5$) and vector ($\Gamma=\gamma_k$, for $k=1,2,3$) meson masses are given by
\begin{eqnarray}
a m_{\Gamma\Gamma} \sim \cosh^{-1} \left[ \frac{C_{\Gamma\Gamma}^{PS}(\tau+1) + C_{\Gamma\Gamma}^{PS}(\tau-1)}{2C_{\Gamma\Gamma}^{PS}(\tau) } \right]\ ,
\end{eqnarray}
and correspond to $a m_\pi$ and $a m_\rho$, respectively.
The bare partially conserved axial current (PCAC) quark mass may be determined from the late-time behavior of
\begin{eqnarray}
a m^{\textrm (bare)}_q \sim \frac{ C_{\gamma_{05}\gamma_5}^{PS}(\tau+1) -C_{\gamma_{05}\gamma_5}^{PS}(\tau-1) }{ 4 C_{\gamma_5\gamma_5}^{PS}(\tau)}.
\end{eqnarray}
The bare pion decay constant is given by
\begin{eqnarray}
a f^{\textrm (bare)}_\pi = \frac{Z_{\gamma_{05}}^P}{am_\pi}\ ,
\end{eqnarray}
and may be determined from the pseudoscalar mass and overlap factors extracted from $C_{\gamma_{05}\gamma_5}^{PS}(\tau)$ and $C_{\gamma_5\gamma_5}^{SS}(\tau)$.
The renormalized PCAC mass and pion decay constant are given by $m_q = ({\cal Z}_A/{\cal Z}_P) m^{\textrm (bare)}_q$ and $f_\pi = {\cal Z}_A f^{\textrm (bare)}_\pi$, respectively.
Note that ${\cal Z}_A$ is renormalization-scale independent, and at one-loop in perturbation theory is given by ${\cal Z}_A = 1-g_0^2 C_F d_A(1)$, where $C_F=3/4$ is the quadratic Casimir for fermions in the fundamental representation of the gauge group $SU(2)$, and $d_A(1)=0.100030(2)$~\cite{Groot:1983ng} (see also~\cite{Martinelli:1982mw} and~\cite{Meyer:1983ds}).

The isovector meson correlation functions described above were estimated using a single measurement per configuration.
The source location was selected at random in order to reduce autocorrelations between measurements.
Each correlator measurement was self-averaged under time reversal, noting that although time reversal is a symmetry of the correlators considered, it is not a symmetry of the background gauge configurations on which they are estimated.
Correlators were finally averaged into $50$ contiguous blocks, resulting in a further reduction of autocorrelations.
Correlated single-exponential least-squares fits were performed within the time interval $[0,T/2]$ in order to extract the leading overlap factors and exponents of the correlation function.
Statistical uncertainties were determined via a bootstrap analysis; systematic uncertainties were estimated by performing scans over all possible fit ranges (with fit windows greater than five time steps), accepting fits with a $\chi^2$ per degree of freedom (d.o.f.) less than unity, and finally taking the standard deviation of the central values obtained for all acceptable fits.
Results from this analysis, including both statistical and systematic uncertainties, are summarized in \Tab{ensembles_measurements}.

\begin{table}
\caption{%
\label{tab:ensembles_measurements}%
Estimated observables and scales on coarse and fine ensembles.~\footnote{The pion decay constant $a f^{\textrm (bare)}_\pi$ differs from that of~\cite{Detmold:2014kba} by a factor of $e^{-am_\pi/2}$, which appears to be due to a mislabeling of the time separation in correlation functions by a single lattice spacing in that work. This discrepancy (e.g., for ensemble $C_{12}$) is a lattice artifact, and vanishes in the continuum limit.}
}
\begin{ruledtabular}
\begin{tabular}{cccccccc}
Label & $\langle \bar W \rangle$ & $\langle \bar\psi \psi \rangle^{\textrm (bare)}$ & $t_0/a^2$ & $a m^{\textrm (bare)}_q$ & $a f^{\textrm (bare)}_\pi$ & $a m_\pi$ & $a m_\rho$ \\
\hline
\input{tab1.tex}
\end{tabular}
\end{ruledtabular}
\end{table}

In addition to correlation functions, and the associated scales extracted from them, we consider the reference scale $t_0$ obtained by the Wilson flow equations
\begin{eqnarray}
\frac{d}{dt} V_\mu(x;t) = -g_0^2 \{ \partial_{x\mu} S_g \} V_\mu(x;t) \ ,\qquad V_\mu(x;0) = U_\mu(x)\ ,
\end{eqnarray}
and the condition $t_0^2 E(t_0) = 0.3$, where $E(t)$ is the symmetric (clover) action density constructed from $V_\mu(x;t)$~\cite{Luscher:2010iy}.
The flow radius at this reference scale corresponds to $\rho_0 =\sqrt{8 t_0}$; throughout, we consider lattice volumes $L \sim 4 \rho_0 -5 \rho_0$.
Wilson flow was performed using a step size of $0.01$, as described in~\cite{Luscher:2010iy}.
Estimates of $t_0$ were determined by a linear interpolation of central values and uncertainties of $t^2 E(t)$ closest to $0.3$ [integrated autocorrelation times required for a reliable estimate of statistical uncertainties were determined using the closest measured values of $t^2E(t)$ to $t_0$].
Results from this analysis, including statistical uncertainties, are provided in \Tab{ensembles_measurements}.

Next, we briefly review the refinement prescription introduced in~\cite{Endres:2015yca}, which provides the basis for the prescription used in this work (as will be discussed in later sections, this refinement prescription will be augmented by a small number of subsequent quenched HMC updates).
First, the coarse link variables are mapped from the coarse lattice with lattice spacing $2a$ to the fine lattice with lattice spacing $a$ following the prescription
\begin{eqnarray}
U_\mu^{f}(x) = \left\{ \begin{array}{ll}
 U_\mu^{c}(x/2) & \textrm{if $x/a$ mod $2 = 0$}\\
 1 & \textrm{otherwise}
\end{array} \right. \ ,
\end{eqnarray}
for each $\mu$, where the superscript labels fine ($f$) and coarse ($c$) link variables, respectively.
Subsequently, the transferred link variables are interpolated into the coarse hypercube bulk following one of a variety of procedures (see, e.g.,~\cite{Luscher:1981zq,Phillips:1986qd,'tHooft1995491}).
In this work, we follow a variation on the method of 't Hooft~\cite{'tHooft1995491}.
We begin by introducing the functions
\begin{eqnarray}
\chi(x) &=& \sum_\mu (\textrm{$x_\mu/a$ mod $2$}) \ , \cr
\chi_\mu(x) &=& \chi(x-x_\mu e_\mu) \ ,\cr
\chi_{\mu\nu}(x) &=& \chi(x-x_\mu e_\mu-x_\nu e_\nu) \ ,
\end{eqnarray}
which map the $p$-cells of the fine lattice ($p=0,\ldots,2$ indicates the number of indices on symbol $\chi$) onto the integers $0, \ldots, 4-p$.
The interpolation is then achieved by sequentially minimizing the actions 
\begin{eqnarray}
S_g^{(d)} = -\frac{\beta}{2}\sum_x \sum_{\mu<\nu} \delta_{d,\chi_{\mu\nu}(x)} W_{\mu\nu}(x)
\end{eqnarray}
with respect to link variables that satisfy $\chi_\mu(x) = d+1$, starting from $d=0$.
At each stage $d=0,1,2$, the interpolation is carried out by successive applications of APE smearing~\cite{Falcioni1985624,Albanese1987163} (appropriately modified) using the smearing factor $0.05$; the smearing is terminated at each stage when the relative change in $S_g^{(d)}$ becomes less than $0.001$\%.
Further details of this procedure can be found in~\cite{Endres:2015yca}.

\begin{table}
\caption{%
\label{tab:match_fits}%
Results obtained from uncorrelated least-squares fits to scale estimates provided in \Tab{ensembles_measurements} using \Eq{fitting_model} as the fit model, with $N_\lambda=2$.
Observables estimated on ensembles $C_4$-$C_{10}$, and $C_{12}$ where included in the fits.
}
\begin{ruledtabular}
\begin{tabular}{cccccccc}
$a^{\Delta_Q} Q$ & $\chi^2/\textrm{d.o.f.}$ & $Q^{(0,0)}$ & $Q^{(1,0)}$ & $Q^{(0,1)}$ & $Q^{(2,0)}$ & $Q^{(0,2)}$& $Q^{(1,1)}$ \\
\hline
\input{tab2.tex}
\end{tabular}
\end{ruledtabular}
\end{table}

\section{Renormalization Group Matching}
\label{sec:rg_matching}

\begin{figure}
\includegraphics[width=\figWidthHalf]{\figdir 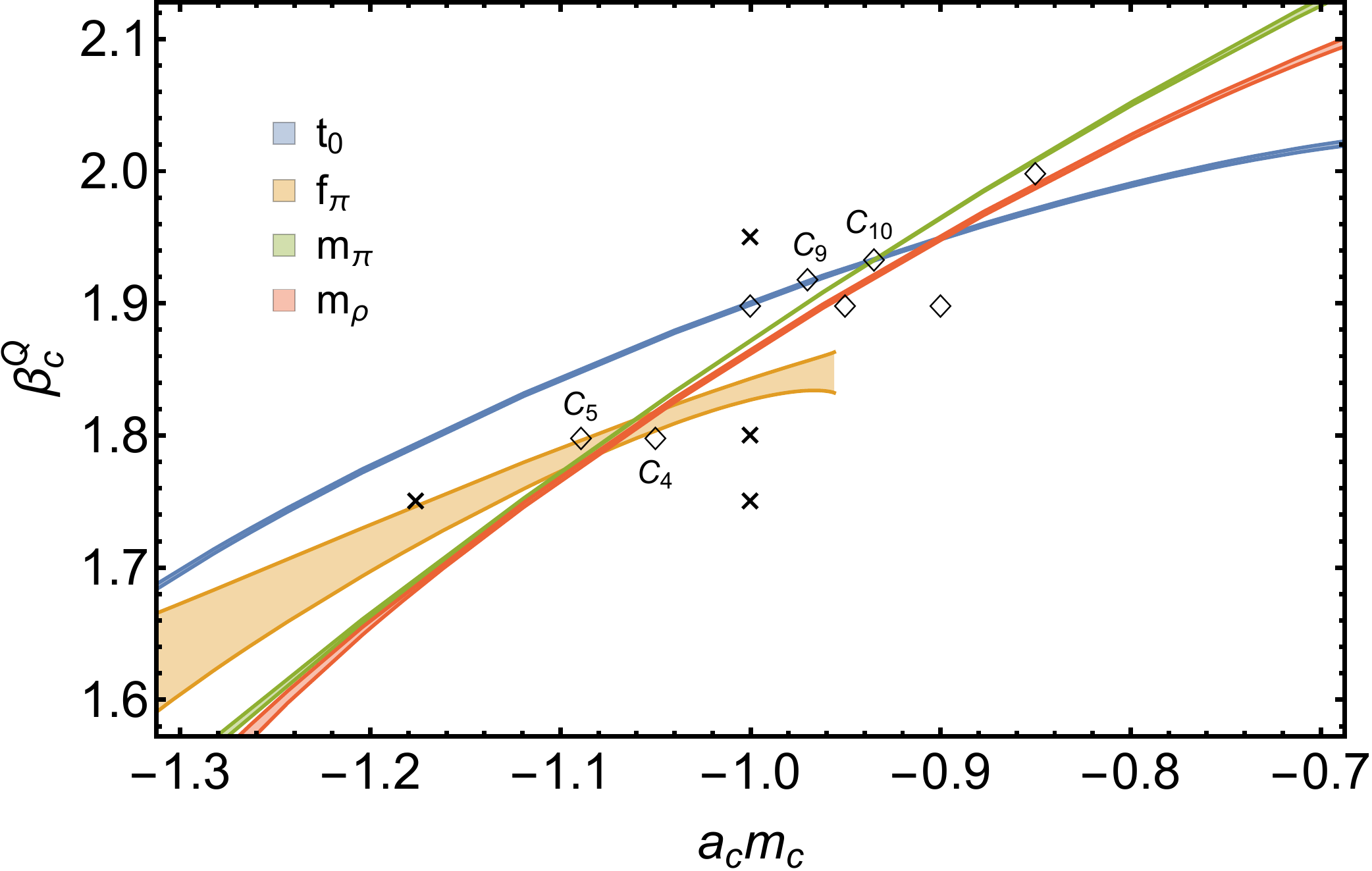}
\includegraphics[width=\figWidthHalf]{\figdir 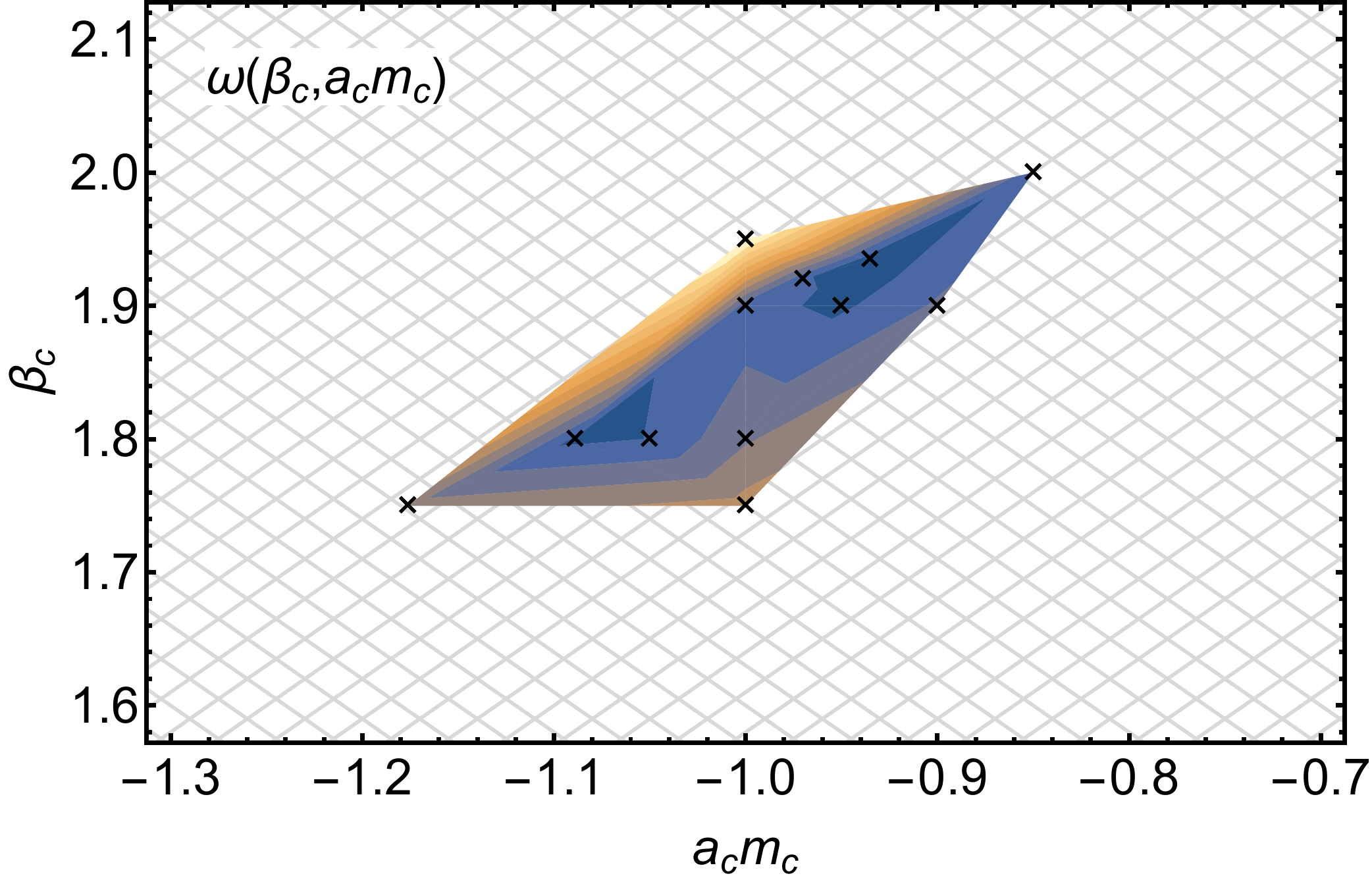}
\caption{\label{fig:coarse_tuning}% 
Left: Numerical determination of $\beta^Q_c$ as a function of $a_c m_c$ for $Q = t_0, f_\pi, m_\pi, m_\rho$, given a fine action corresponding to $\beta_f=2.2$ and $a_f m_f=-0.7200$.
Diamonds ($\diamond$) indicate the coarse ensembles included in the fits (i.e., $C_4$ - $C_{10}$, $C_{12}$), whereas crosses ($\times$) indicate excluded coarse ensembles.
Candidate RG matched coarse ensembles are explicitly labeled, whereas all other ensembles can be identified from \Tab{ensembles_generation}.
Right: $\omega(\beta_c,a_c,m_c)$ obtained for the same fine action and set of scales; blue regions correspond to minima, and the hatched region is undetermined given the available data.
}
\end{figure}

For a given scale-setting quantity, $Q$, and a fixed set of fine couplings $(\beta_f,a_f m_f)$, we define a function $\beta_c^Q(a_c m_c)$ by demanding
\begin{eqnarray}
a_f^{\Delta_Q} Q(\beta_f,a_f m_f) =a_c^{\Delta_Q} Q(\beta_c,a_c m_c)\ ,
\label{eq:obs_matching}
\end{eqnarray}
where $\Delta_Q$ is the mass dimension of the scale, and for simplicity we consider the case where $a_c = 2 a_f$.
The coarse action can then be matched to the fine action by demanding
\begin{eqnarray}
\beta_c^Q(a_c m_c) = \beta_c^{Q^\prime}(a_c m_c)
\end{eqnarray}
for any pair of scales $Q$ and $Q^\prime$ (more generally, the requisite number of scales needed to define a matching condition equals the number of coarse couplings).
For this study, we consider the scales $Q = t_0$, $f_\pi$, $m_\pi$, and $m_\rho$ for the matching, which have mass dimensions $\Delta_Q = -2$, $1$, $1$, and $1$, respectively.
The functions $\beta^Q_c(a_c m_c)$ were obtained by first fitting estimates of $Q$ at various coarse couplings to the functional form
\begin{eqnarray}
a^{\Delta_Q} Q(\beta,a m) = \sum_{i=0}^{N_\lambda} \sum_{j=0}^{N_\lambda-i} \beta^i Q^{(i,j)} (am)^j\ ,
\label{eq:fitting_model}
\end{eqnarray}
treating the $Q^{(i,j)}$ as fit parameters.
Note that this functional form is the simplest that one can consider and does not account for possible logarithmic and other nonpolynomial dependence of the couplings that may be present.
Results of these fits, including the goodness of each fit, are provided in \Tab{match_fits} for $N_\lambda=2$.
These fits were performed using observables estimated on the ensembles $C_4$-$C_{10}$ and $C_{12}$.
Results for $\beta^Q_c(a_c m_c)$, obtained from the fitted data and observable estimates on the fine ensemble, $F$, are shown in \Fig{coarse_tuning} (left).
Error bands include statistical uncertainties associated with the fits, along with statistical uncertainties associated with the observable estimates on the fine ensemble.
Intersecting curves obtained with this procedure yield different, but equally valid, matching conditions.
In practice, however, the quality of the fits and the mutual orthogonality of certain observables makes some combinations better discriminants than others.
Values of coarse couplings corresponding to $C_x$ ($x=1,\ldots,12$) are shown in \Fig{coarse_tuning} (left); the subset of ensembles which have been used for the fit to \Eq{fitting_model} are also indicated in the figure.

Differences in the matching conditions for various pairs of observables result from discretization artifacts, which have been omitted from \Eq{obs_matching}, and model dependence arising from the choice of fit function.
Finite volume effects are expected to be negligible for the couplings considered in the analysis based on the findings of earlier studies, which use comparably tuned couplings~\cite{Detmold:2014kba}.
In particular, for all the fitted data, ensembles satisfy to $m_\pi L \gtrsim 9$.
The curves obtained for each observable in \Fig{coarse_tuning} (left) indicate that $m_\pi$ and $m_\rho$ are fairly correlated, tracking similar paths in the space of couplings.
The reference scale $t_0$, on the other hand, appears to have coupling dependence that is orthogonal to $m_\pi$ and $m_\rho$, which is presumably due to the fact that this scale is derived from a purely gluonic observable.
Although $f_\pi$ appears orthogonal to the other observables, and compared to $t_0$, has an inconsistent intersection with curves for $m_\pi$ and $m_\rho$, this quantity was renormalized using one-loop perturbation theory, and therefore may possess significant systematic errors.
Taking the difference between tree-level and one-loop results as a measure of the uncertainty in ${\cal Z}_A$, we assign a 5\% systematic uncertainty on $\beta_c^{f_\pi}(a_c m_c)$, which is not displayed in the figure.
It must be noted that the fit for $t_0$ is of rather low quality;
the fit quality presumably improves by considering more general functional forms and correspondingly more data.
Increasing the order of the fit to $N_\lambda=3$ and using all data except $C_{11}$, for example, yields a better quality fit, but qualitatively identical results, suggesting that the model dependence of the results is mild.
Given that the goal here is to provide an estimate of the RG matched couplings from which to begin rethermalization studies, any errors in tuning are corrected by rethermalization and therefore the tuning need not be performed with high precision.
In practice, the rethermalization times can also be used to iteratively refine the parameter space search, as we will see below.

We may construct a second, model-independent measure for the matching, by considering the quantity
\begin{eqnarray}
\omega(\beta_c,a_c m_c) = 2 \sum_Q \frac{ |a_f^{\Delta_Q} Q(\beta_f,a_f m_f) -a_c^{\Delta_Q} Q(\beta_c,a_c m_c) | }{ | a_f^{\Delta_Q} Q(\beta_f,a_f m_f) + a_c^{\Delta_Q} Q(\beta_c,a_c m_c)|  } \ ,
\label{eq:obs_weight}
\end{eqnarray}
as a function of the coarse couplings, given a fixed fine coupling.
In the limit that $a_f\to0$, with $a_c/a_f$ held fixed, this measure will presumably yield a single minimum in the coarse coupling plane (assuming only one fixed point corresponding to a continuum limit); this minimum would correspond to vanishing $\omega(\beta_c,a_c m_c)$, up to discretization artifacts, and thus identify a $Q$-independent matching condition.
Although there is some ambiguity in how the operators should be weighted (for simplicity we only consider a uniform weight, and ignore correlations between scales), this choice is expected to become irrelevant in the continuum limit.~\footnote{It is possible the RG matching functional could have multiple minima, although we do not pursue this possibility here.}
A plot of $\omega(\beta_c,a_c m_c)$ is shown in \Fig{coarse_tuning} (right), and appears largely consistent with the previous, model-dependent, analysis in the sense that minima occur in regions of multiple crossings in \Fig{coarse_tuning} (left).

While the matching procedures discussed above are imperfect, they suggest that the actions corresponding to ensembles $C_4$, $C_5$, $C_9$, and $C_{10}$ are good candidate RG matched coarse actions for the purpose of achieving rapid rethermalization times.
We will consider these particular choices in greater detail in the next sections.
Given the perturbative renormalization for $f_\pi$, however, the cases $C_4$ and $C_5$ should be considered with care. 
As the continuum limit is approached, the ambiguities in the choice of the scales involved in matching will diminish and a more precise prediction is expected to emerge.
Eventually, the RG will become dominated by perturbative scales for which analytic predictions for the matched action can be utilized.

\section{Dirac Spectrum}
\label{sec:dirac_spectrum}

The spectra of the various Dirac operators are interesting to investigate as they provide another handle on the RG matching of the coarse and fine actions.
Furthermore, dynamical fermions generate additional forces (governed by the Dirac operator) which drive the HMC evolution in QCD simulations.
The effects of such forces have not yet been considered in the multiscale thermalization studies of pure Yang-Mills theory.
In order to construct an efficient multiscale thermalization algorithm, a prolongation scheme that deforms the initial coarse Dirac spectrum into a close approximation to the fine distribution is desirable.
{\it A priori}, it is unclear that the gauge field interpolation introduced in~\cite{Endres:2015yca} will fill this role.
Indeed, as will be demonstrated in the next section, the short-distance defects from this interpolation produce a density of spurious low modes of the Dirac operator and additional steps must be included in the interpolation scheme to remove them.

For two-color QCD, the Wilson-Dirac operator satisfies the properties $\gamma_5^\dagger D_w \gamma_5 = D_w^\dagger$ and $C^\dagger D_w C = D_w^*$, where $C$ is the charge conjugation operator.~\footnote{Recall that $C=-C^\Transpose$, $C^\dagger C=1$, $C^\dagger \gamma_\mu C = \gamma_\mu^\Transpose$, $C^\dagger \gamma_5 C = \gamma_5^\Transpose=\gamma_5^*$, and $C^\dagger T^a C = - (T^a)^\Transpose$ where $T^a$ are the generators of a real or pseudoreal representation of the gauge group [for $SU(2)$, all representations are pseudoreal].}
If $\psi$ is an eigenvector of the positive definite operator $M= \left(D_w^\dagger D_w\right)^{1/2}$ with eigenvalue $\lambda$, one can show from the relations above that $C \psi^*$ is also an eigenvector of $M$ with eigenvalue $\lambda$.
Note that $\psi$ and $C \psi^*$ are orthogonal eigenvectors of M (i.e., they are independent) due to the antisymmetry of $C$.
For this study, the lowest 200 (400) unique eigenvalues of $M$ were determined up to a residual norm tolerance of $10^{-9}$ on $n_\textrm{conf}=13$ decorrelated configurations within each coarse (fine) ensemble.
Accounting for the twofold degeneracy of eigenvalues, the eigenvalue density is approximated by
\begin{eqnarray}
\rho_n(\Delta\lambda) = \frac{1}{4V n_\textrm{conf} \Delta\lambda} \int_{n\Delta\lambda}^{(n+1)\Delta\lambda} d\lambda^\prime \sum_{i} \delta(\lambda^\prime - \lambda_i)
\end{eqnarray}
where $i=0,\ldots, 4Vn_\textrm{conf}-1$ is a collective index, which labels the unique eigenvalues of $M$ on all configurations.
Defined in this way, the approximate eigenvalue density is normalized such that 
\begin{eqnarray}
\Delta\lambda \sum_{n\ge0} \rho_n(\Delta\lambda) = 1\ ;
\end{eqnarray}
in this work, we take bins of size $\Delta\lambda = 1/(200 \sqrt{t_0})$.

\begin{figure}
\includegraphics[width=\figWidthHalf]{\figdir 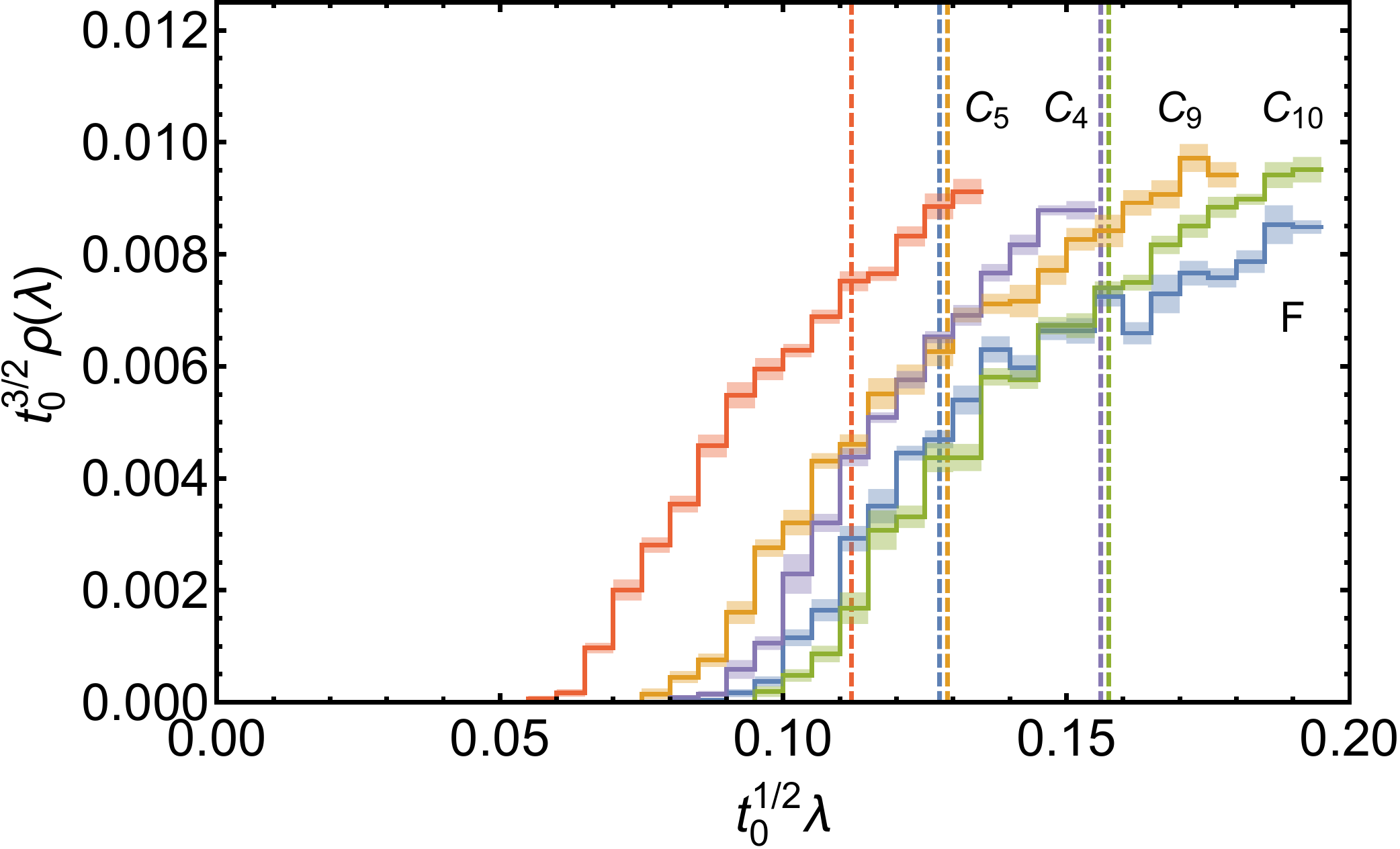}
\includegraphics[width=\figWidthHalf]{\figdir 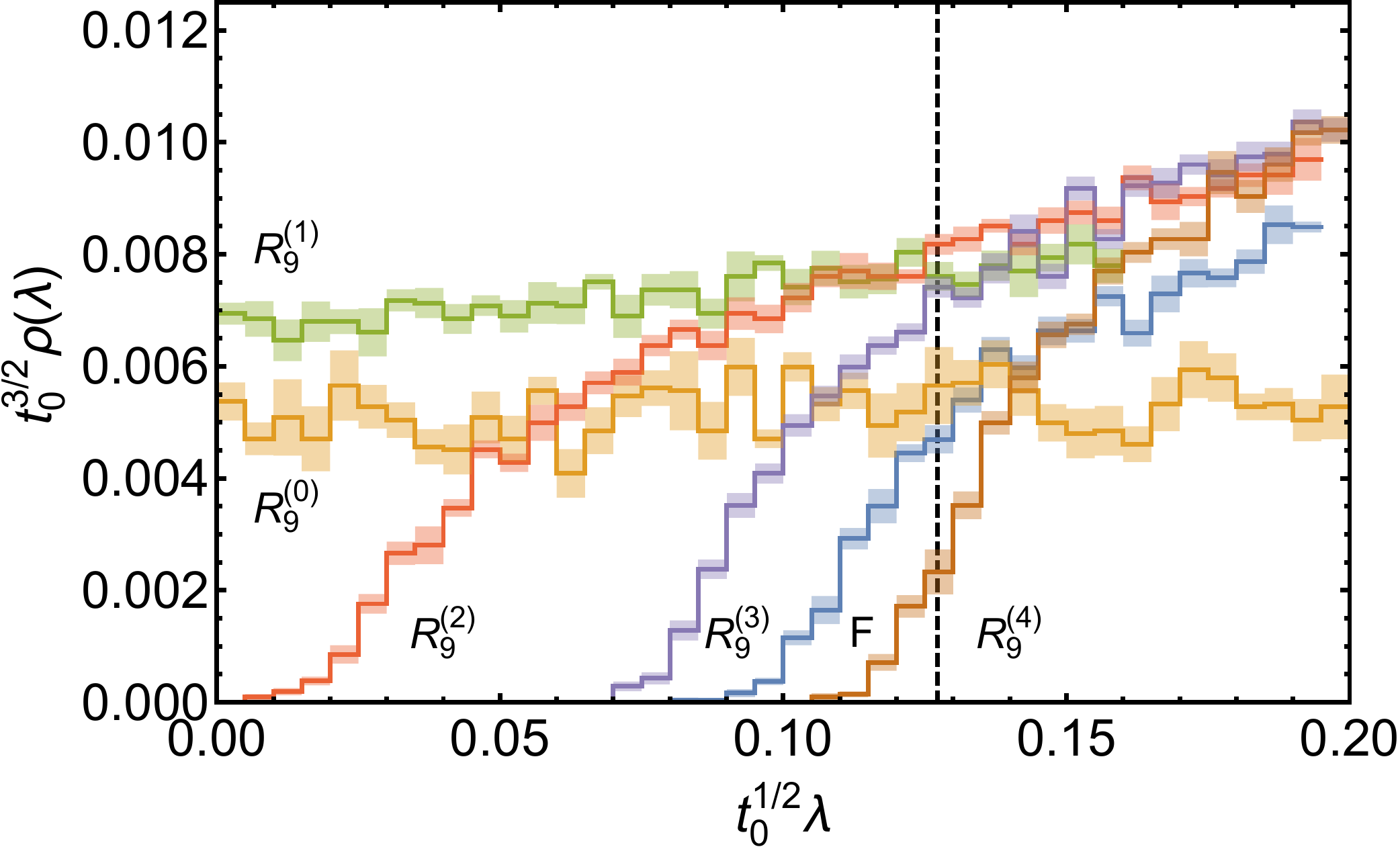}
\caption{\label{fig:dirac_eig}% 
Left: Wilson-Dirac spectrum measured on a variety of thermalized ensembles.
Vertical lines correspond to $t_0^{1/2} m^{(bare)}_q$ measured on $C_5$, $F$, $C_9$, $C_4$ and $C_{10}$ (ordered from left to right).
Right: Wilson-Dirac spectrum measured on ensemble $F$, and on an ensemble that has been refined and evolved using a quenched action.
The vertical band corresponds to $t_0^{1/2} m^{(bare)}_q$ measured on the fine ensemble, $F$.
}
\end{figure}

Plots of the binned eigenvalue density are shown in \Fig{dirac_eig} (left) for the fine ensemble, $F$, and the candidate RG matched coarse ensembles: $C_4$, $C_5$, $C_9$, and $C_{10}$.
The ensemble $C_{10}$ corresponds to simultaneous matching of the scales $t_0$ and $m_\pi$, and appears to have a consistent eigenvalue distribution with $F$.
On the other hand, the ensembles $C_4$, $C_5$ and $C_9$ have eigenvalue distributions which disagree with $F$, despite evidence from the scale matching studies that suggest a potentially good match.
Also shown in \Fig{dirac_eig} (left) is the bare PCAC mass measured on each ensemble, which in all cases is comparable in size to the spectral gap.
Although $C_9$ yields the best agreement with $F$ in terms of the bare PCAC mass, the validity of this matching is diminished by the fact that there are potentially significant and unaccounted for corrections due to the renormalization of this quantity.

\section{Thermalization}
\label{sec:thermalization}

For each coarse ensemble $C_x$ ($x=1,\ldots,12$) listed in \Tab{ensembles_generation}, a small but decorrelated subset of configurations of size $n_\textrm{conf}=13$ were prolongated using the interpolation prescription outlined in \Sec{ensembles}.
We indicate these prolongated ensembles by the label $R_x^{(0)}$.
Before considering gauge evolution, and the rate at which the prolongated configurations return to equilibrium under HMC, we first consider the spectral properties of the Dirac operator.
The Dirac spectrum associated with the prolongated ensembles was determined using the same bare mass parameter as that of the fine ensemble $F$.
In \Fig{dirac_eig} (right), we show the Dirac spectrum obtained for ensemble $F$ and an exemplary refined ensemble, namely, $R_9^{(0)}$.
As noted earlier, in the former case, the spectrum shows a clear gap of the order $am^{(bare)}_q$.
In the latter case, however, we find that the spectrum is uniformly distributed (at the same resolution $\Delta\lambda$ used for $F$) over the entire range of eigenvalues considered.

The presence of near-zero eigenvalues for $R_9^{(0)}$ imply large fermion forces at the initial stages of the rethermalization, which demand extremely small HMC step sizes for stable gauge evolution.
For typical integrator step sizes, such as those described in \Tab{ensembles_generation}, such large forces can drive the prolongated configurations toward a disordered state (signaling an instability in the HMC algorithm), thus destroying the topology and long-distance correlations we had set out to preserve.
To cure this problem, we consider an initial quenched evolution of the prolongated ensembles (using the same evolution parameters as ensemble $F$, but with bare mass $m_0=\infty$).
We indicate prolongated configurations that have undergone $\tau$ trajectories of quenched evolution by the label $R_x^{(\tau)}$.
In \Fig{dirac_eig} (right), we plot the Dirac spectrum for a particular refined and quenched-evolved ensemble at times $\tau=1,2,3, 4$, and see that after only a few iterations, the near-zero modes are removed and the expected shape of the eigenvalue distribution is qualitatively restored.

In light of the fact that a short quenched evolution of the gauge fields can only impact the short-distance properties of the configuration, and the fact that the observed near-zero modes are removed after only a few trajectories, such modes are presumably associated with ultraviolet (UV) aspects of the configuration.
To support this understanding, we consider the thermalization behavior of a particular refined and quench-evolved configuration as a function of the quenched evolution time, and find that for long-distance observables, the thermalization behavior is insensitive to the quenched evolution time at early times (particularly $R_9^{(2)}$ and $R_9^{(3)}$).
For extended quenched evolution times, this is no longer expected to be the case; for this reason, we henceforth chose to incorporate two quenched evolution steps (i.e., the minimal number needed to eliminate instabilities of the subsequent dynamical evolution) into the definition of the prolongation procedure.

\begin{figure}
\includegraphics[width=\figWidthHalf]{\figdir 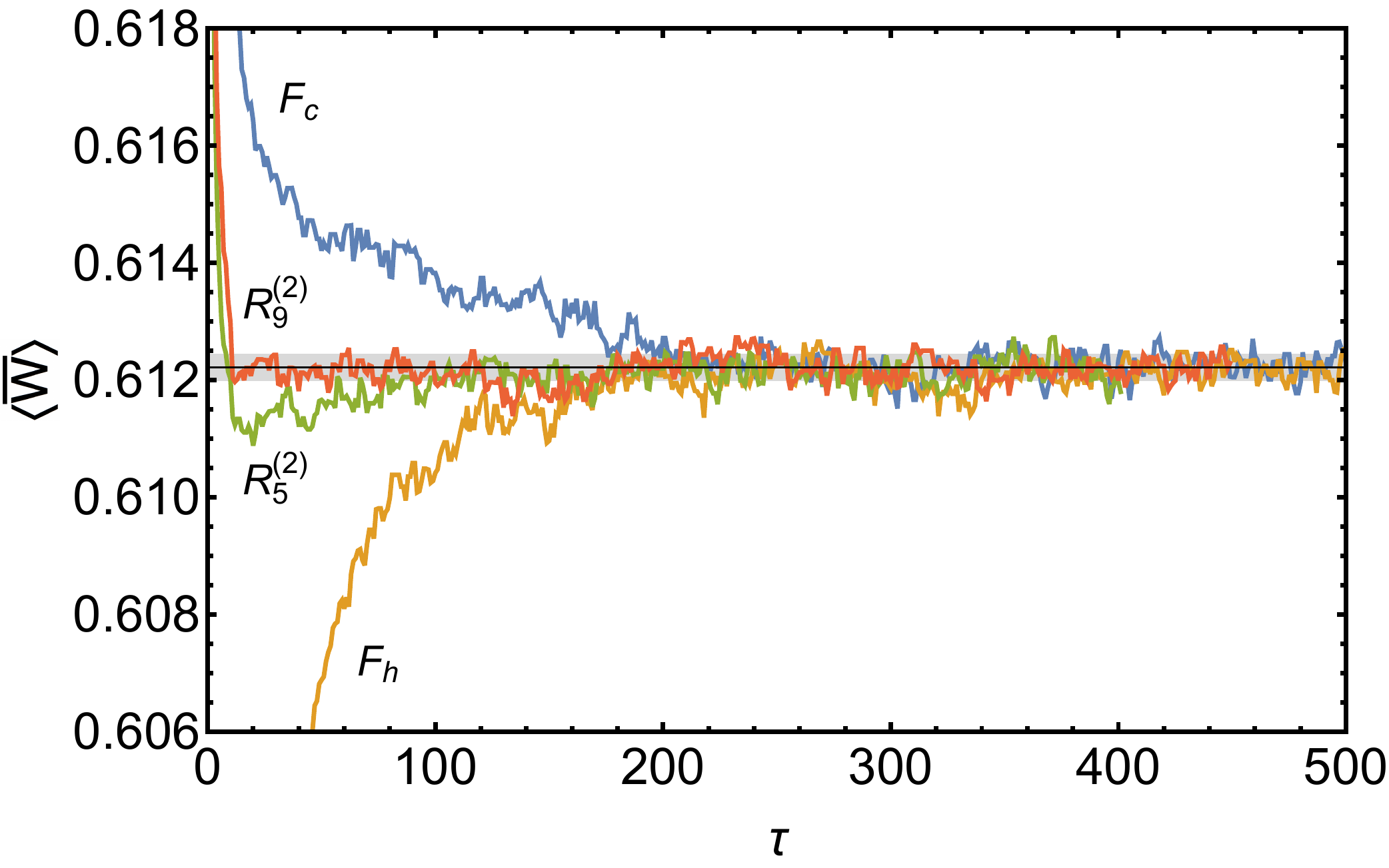} 
\includegraphics[width=\figWidthHalf]{\figdir 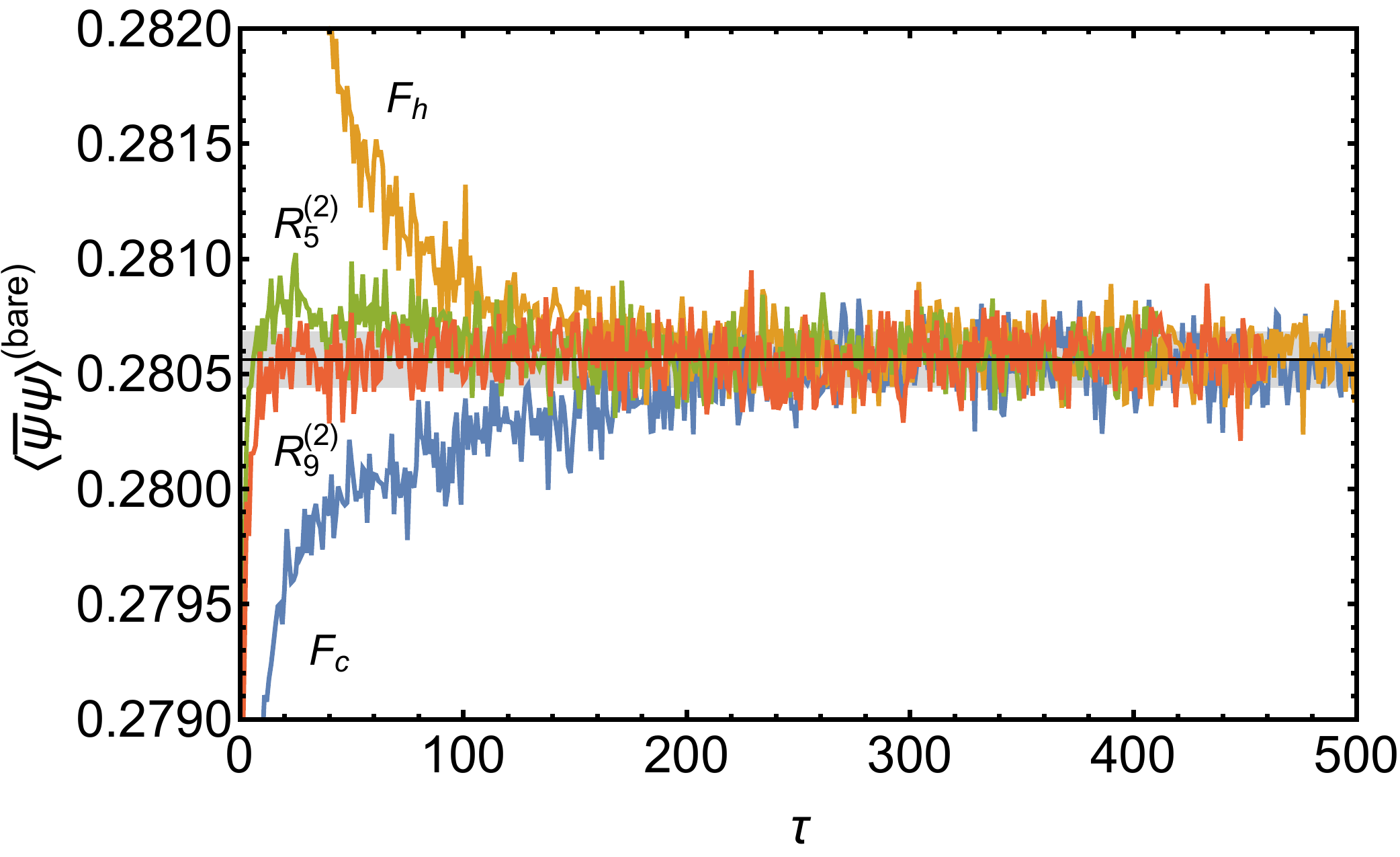} \\
\vspace{12pt}
\includegraphics[width=\figWidthHalf]{\figdir 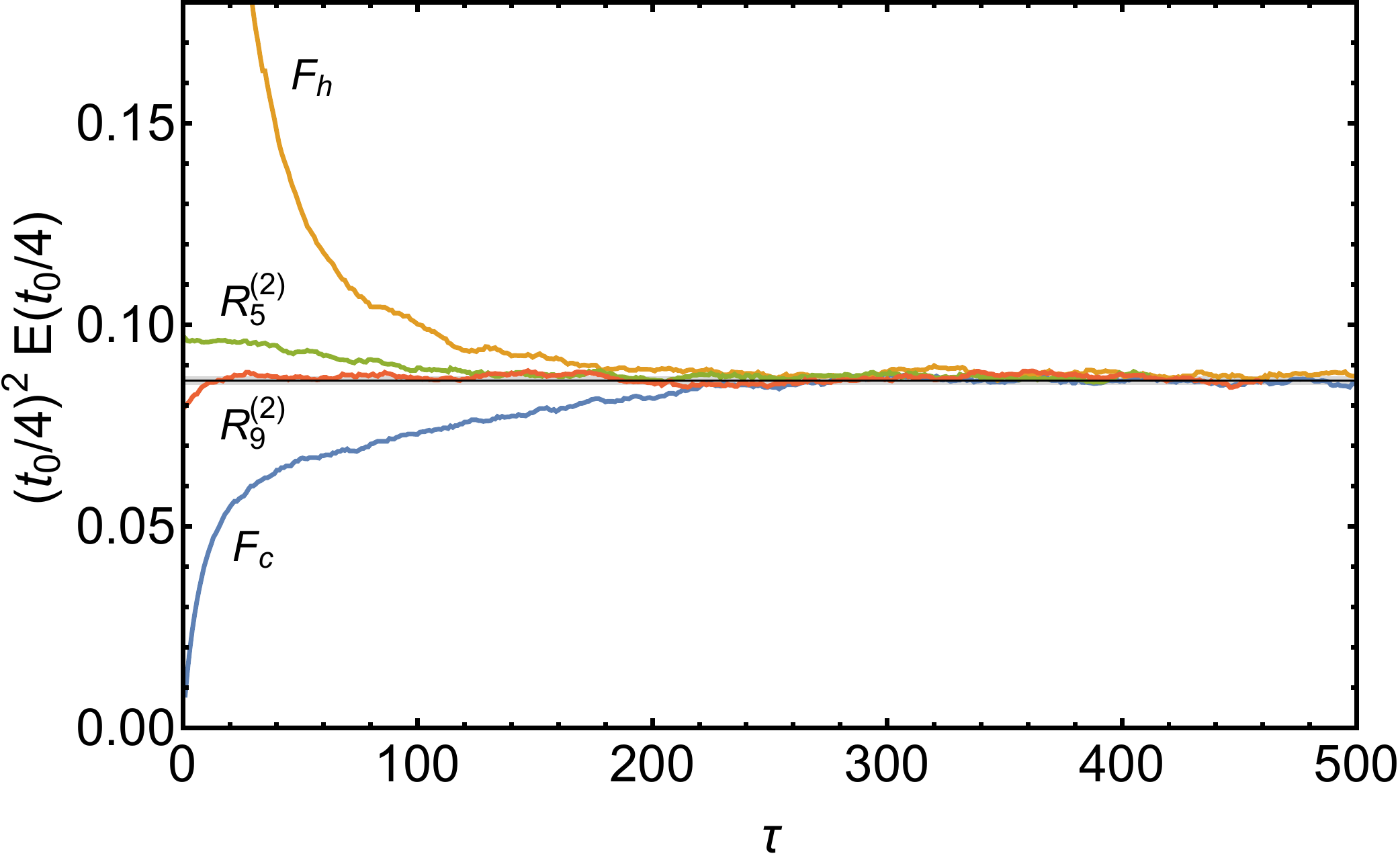} 
\includegraphics[width=\figWidthHalf]{\figdir 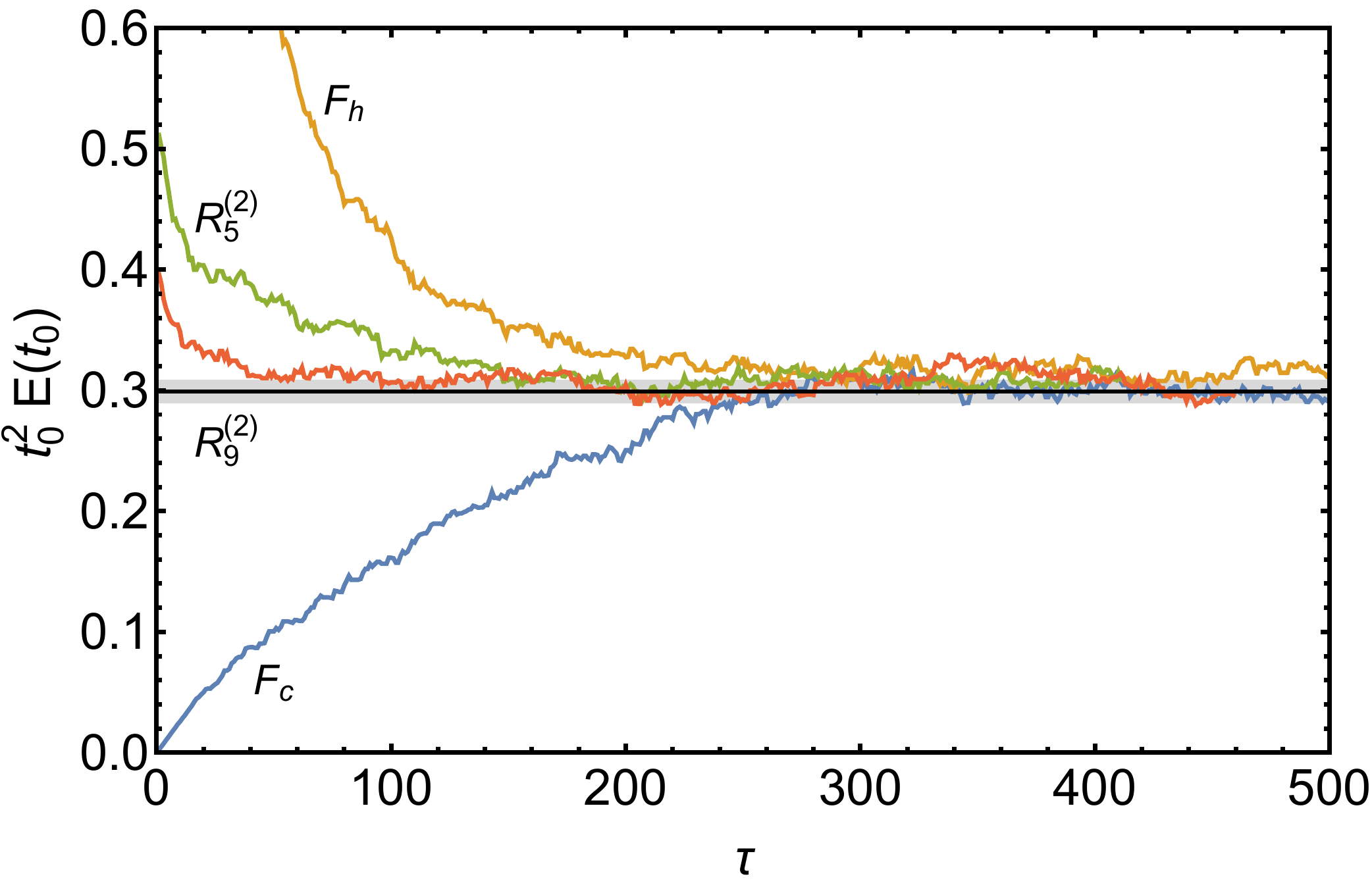} \\
\vspace{12pt}
\includegraphics[width=\figWidthHalf]{\figdir 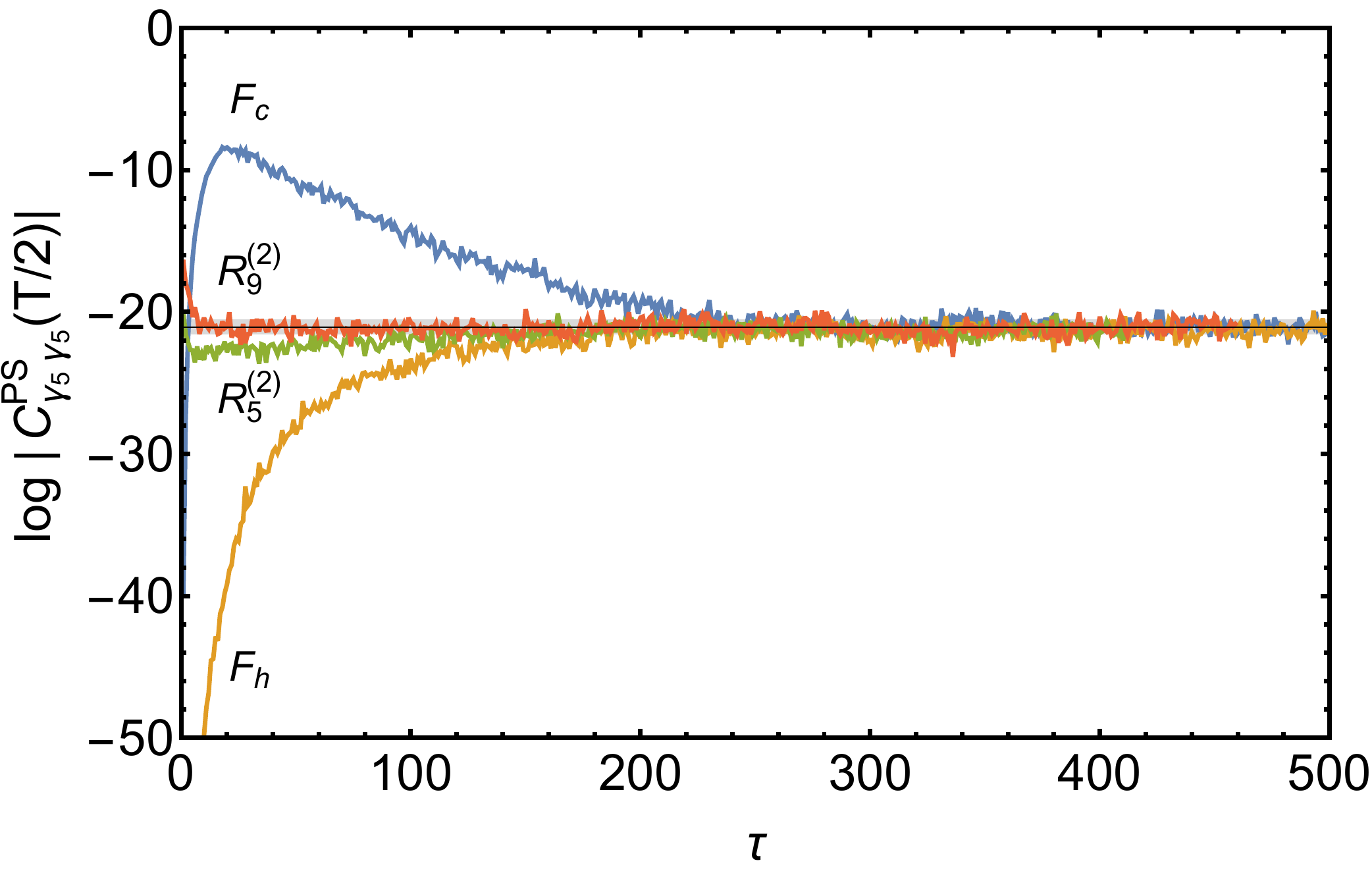}
\includegraphics[width=\figWidthHalf]{\figdir 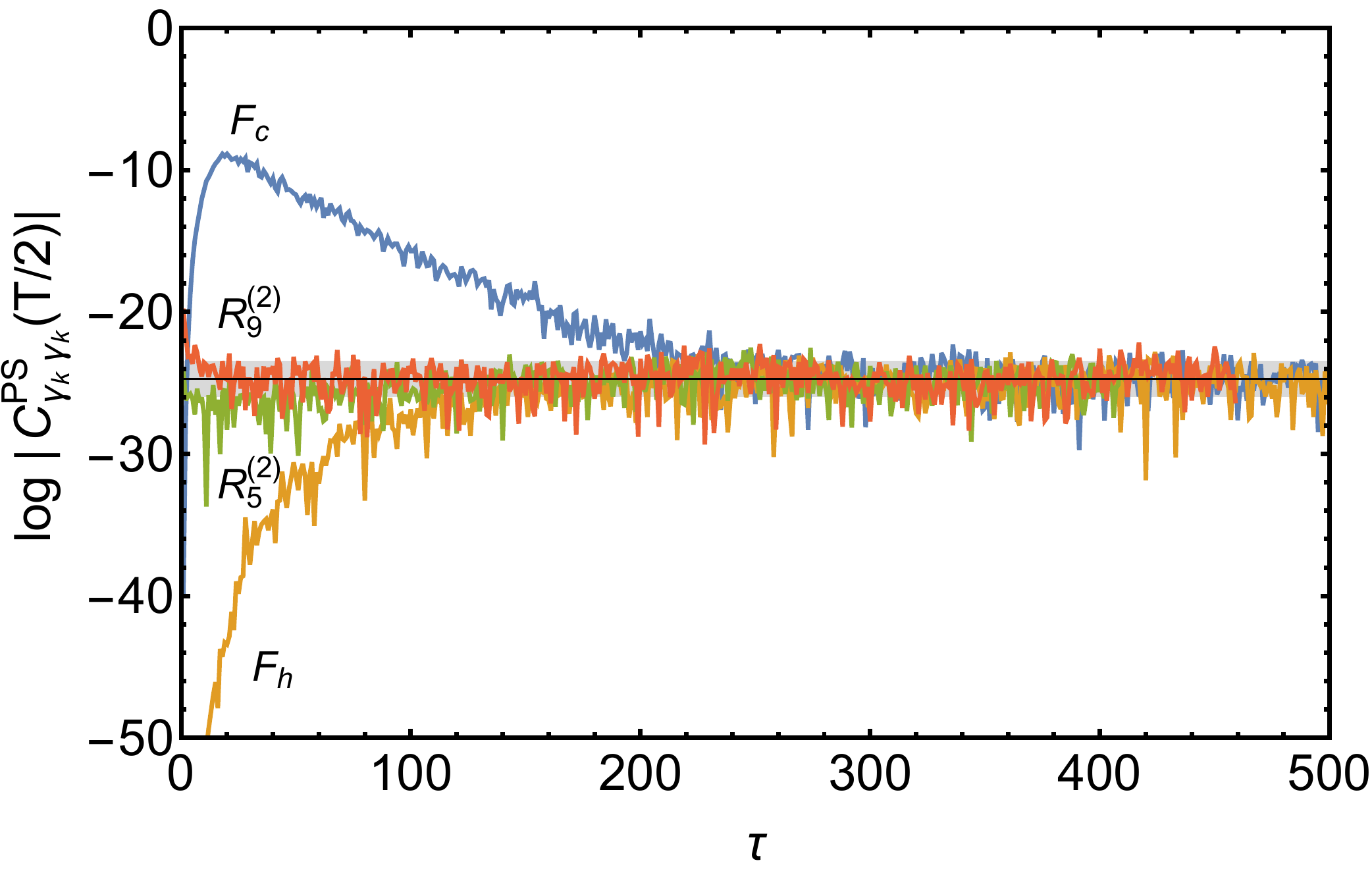}
\caption{\label{fig:therm_curves}% 
Thermalization curves for the plaquette (top, left), chiral condensate (top, right), action density at flow times $t_0/4$ (center, left) and $t_0$ (center, right), and pion (bottom, left) and rho (bottom, right) correlation functions evaluated at $T/2$.
}
\end{figure}

\begin{figure}
\includegraphics[width=\figWidthHalf]{\figdir 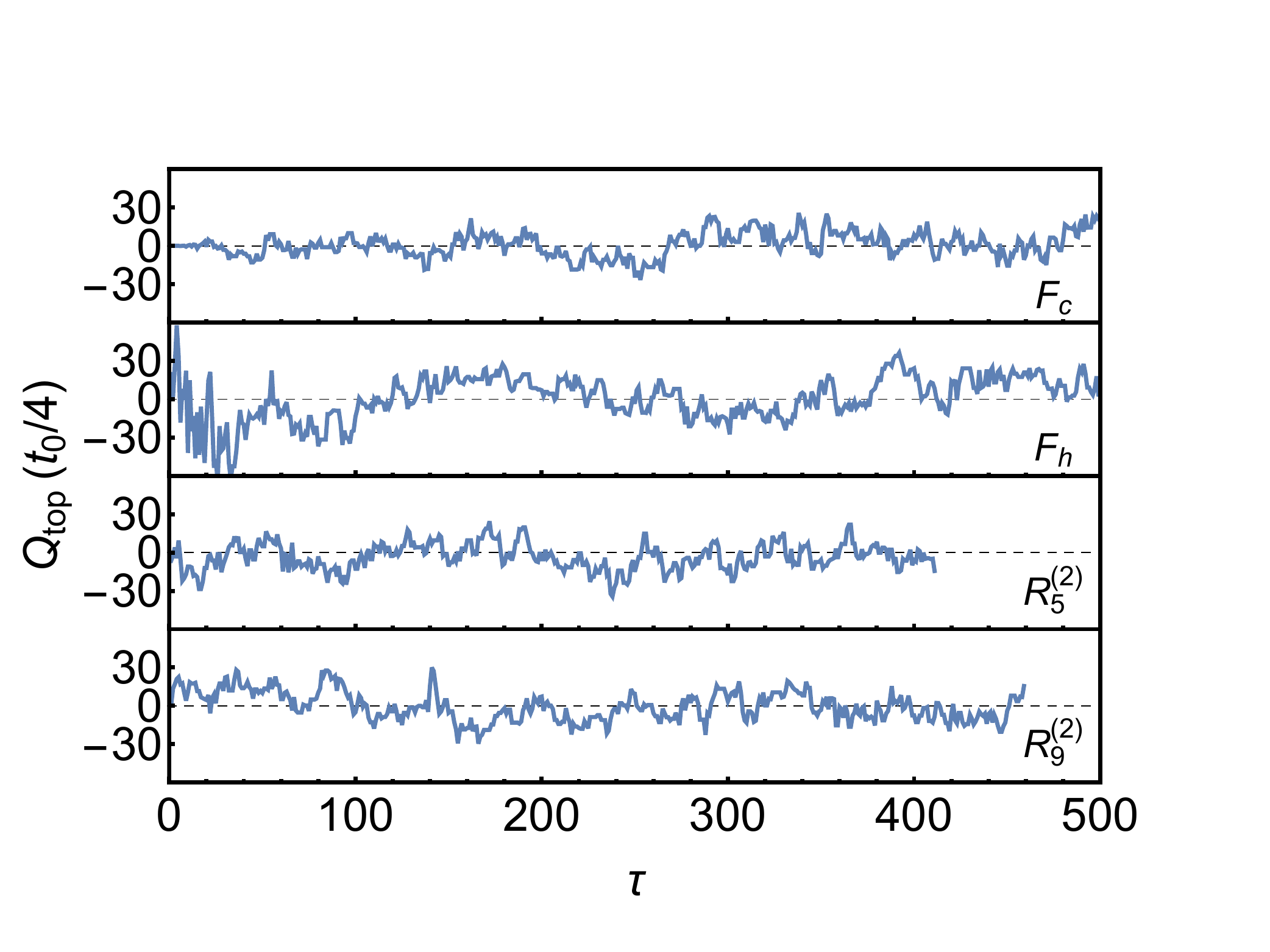} 
\includegraphics[width=\figWidthHalf]{\figdir 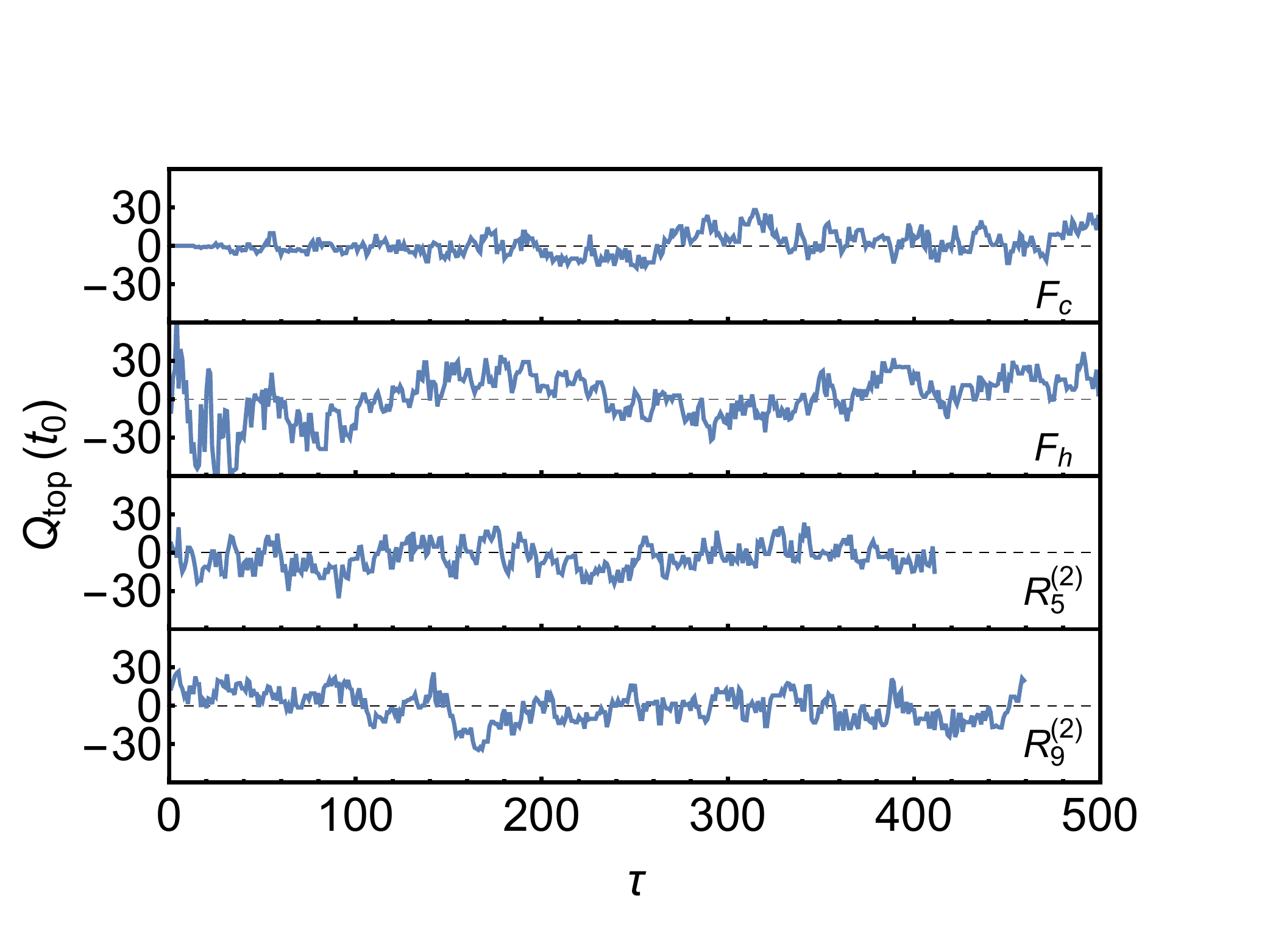} \\
\caption{\label{fig:qtop_curves}% 
Three-loop improved definition of the topological charge operator~\cite{deForcrand:1997esx} evaluated at flow times $t_0/4$ (left) and $t_0$ (right).
}
\end{figure}

For numerical expediency, we consider the rethermalization behavior of {\it single} configurations, drawn from each $R_x^{(2)}$ ($x=1,\ldots,12$), as a function of thermalization time.
In \Fig{therm_curves} we show the thermalization curves for two exemplary cases, namely $R_5^{(2)}$ and $R_9^{(2)}$, for a variety of short- and long-distance observables: $\langle \bar W \rangle$, $\langle \bar \psi \psi \rangle^{\rm (bare)}$, $(t_0/4)^2 E(t_0/4)$, $t_0^2 E(t_0)$, $\log |C_{\gamma_5\gamma_5}^{PS}(T/2)|$, and $\log |C_{\gamma_k\gamma_k}^{PS}(T/2)|$ (fixed $k$).
For reference, we also show thermalization curves for both hot ($F_h$) and cold ($F_c$) starts, corresponding to configurations drawn from a thermalized ensemble in the limits $(\beta=0,m_0=\infty)$ and $(\beta=\infty,m_0=\infty)$, respectively.
Estimates of each observable, determined from the equilibrated fine ensemble described in \Tab{ensembles_generation}, are also indicated, along with the size of fluctuations based on the variance of its distribution.
It should be noted that for the choice of fine couplings considered in this work, the topological charge fluctuates frequently, as illustrated in \Fig{qtop_curves}.

From a theoretical standpoint, all observables (as well as their fluctuations) have a thermalization time dependence given by
\begin{eqnarray}
\langle \calO \rangle_\tau = \langle \calO \rangle + \sum_{n=1} z_n(\calO,\calP) e^{-E_n \tau}\ ,
\end{eqnarray}
where the $\tau_n = 1/E_n$ determine the time scale of the fine evolution, and $\calP$ represents the distribution from which the initial gauge configuration was drawn.
The overlap factors have a separable form, given by
\begin{eqnarray}
z_n(\calO,\calP) = \langle \calO | \chi_n \rangle \langle \tilde \chi_n | \calP \rangle\ ,
\end{eqnarray}
where  $|\chi_n\rangle$ and $\langle \tilde \chi_n|$ are the left and right eigenvectors of the transition matrix $\calM$, which defines the Markov process (in this case HMC); these eigenvectors satisfy $\langle \tilde \chi_n |\calM|\chi_m \rangle = e^{-E_n} \delta_{nm}$  (see, e.g.,~\cite{Schaefer:2010hu}).
Note that $|\chi_n \rangle$ and $\langle \tilde \chi_n|$ are {\it algorithm dependent}, and our objective is to find a distribution $\calP$ with minimal overlap onto the lowest such modes for a given algorithm.
Our underlying assumption is that such a distribution can be approximately constructed by prolongating coarse gauge fields, which have been generated using an RG matched coarse action.
Ideally, the differences between this constructed distribution and the thermalized distribution are dominated by the fast modes of $\calM$.

\begin{figure}
\includegraphics[width=\figWidthHalf]{\figdir 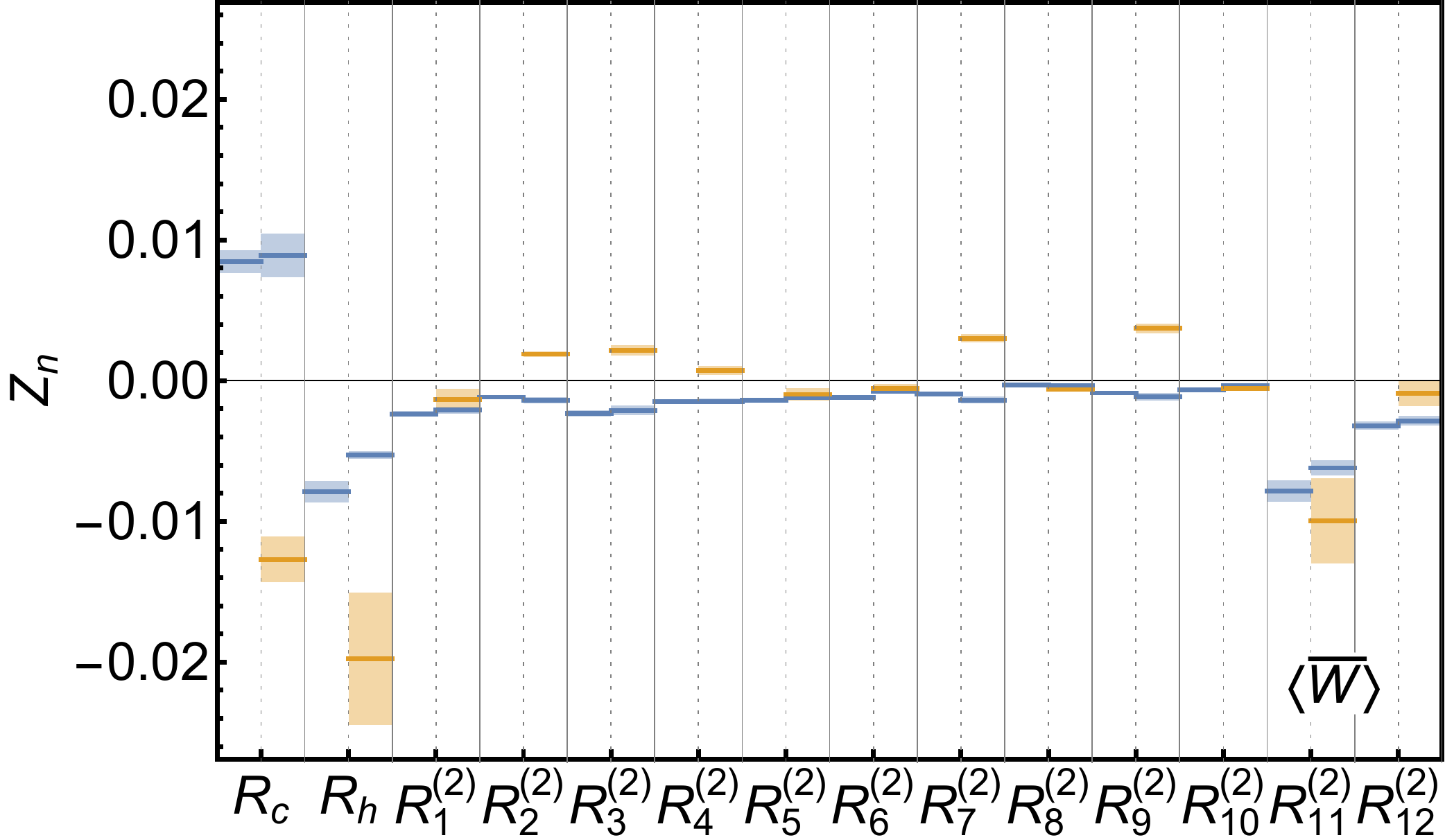} 
\includegraphics[width=\figWidthHalf]{\figdir 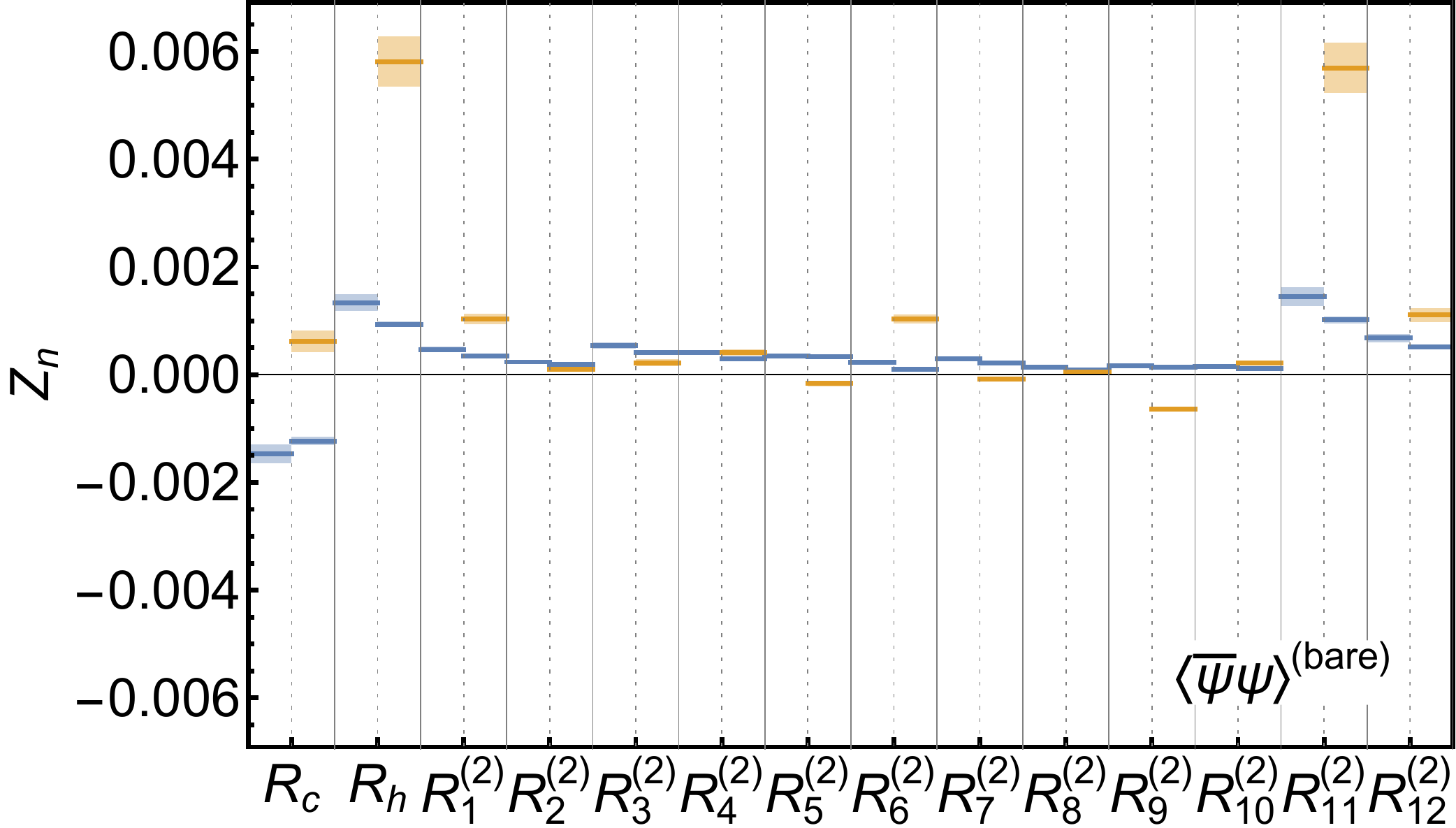} \\
\vspace{12pt}
\includegraphics[width=\figWidthHalf]{\figdir 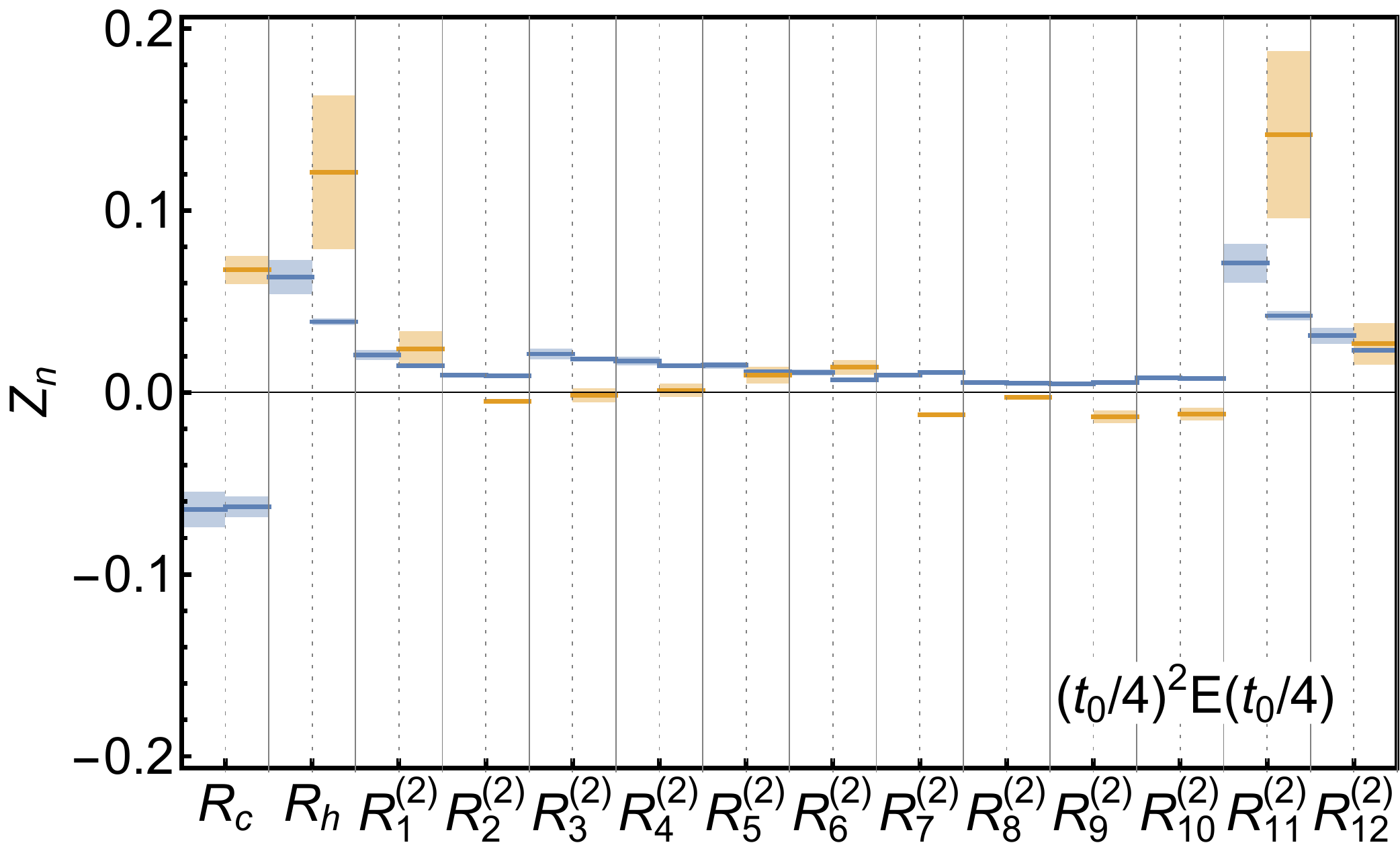}
\includegraphics[width=\figWidthHalf]{\figdir 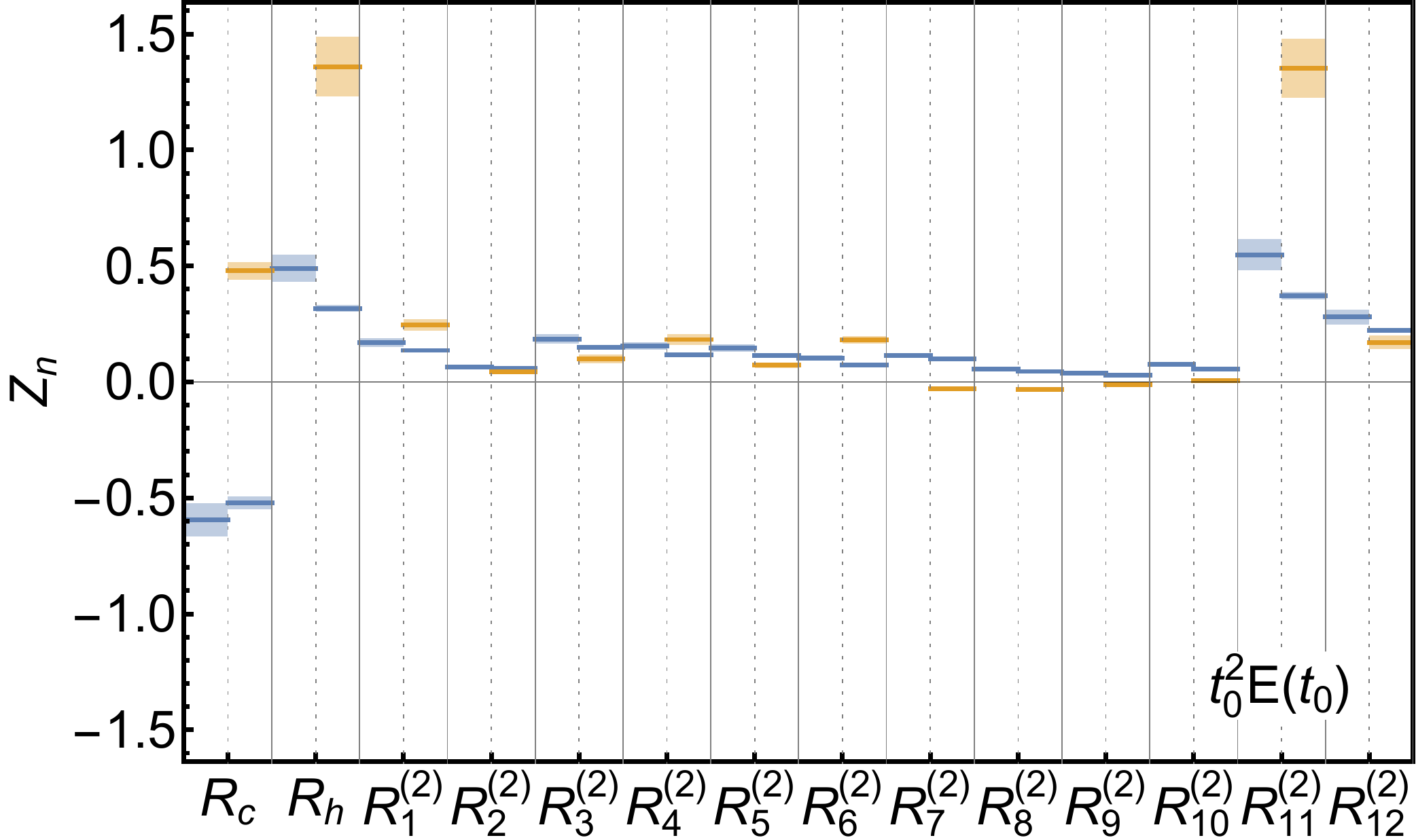} \\ 
\vspace{12pt}
\includegraphics[width=\figWidthHalf]{\figdir 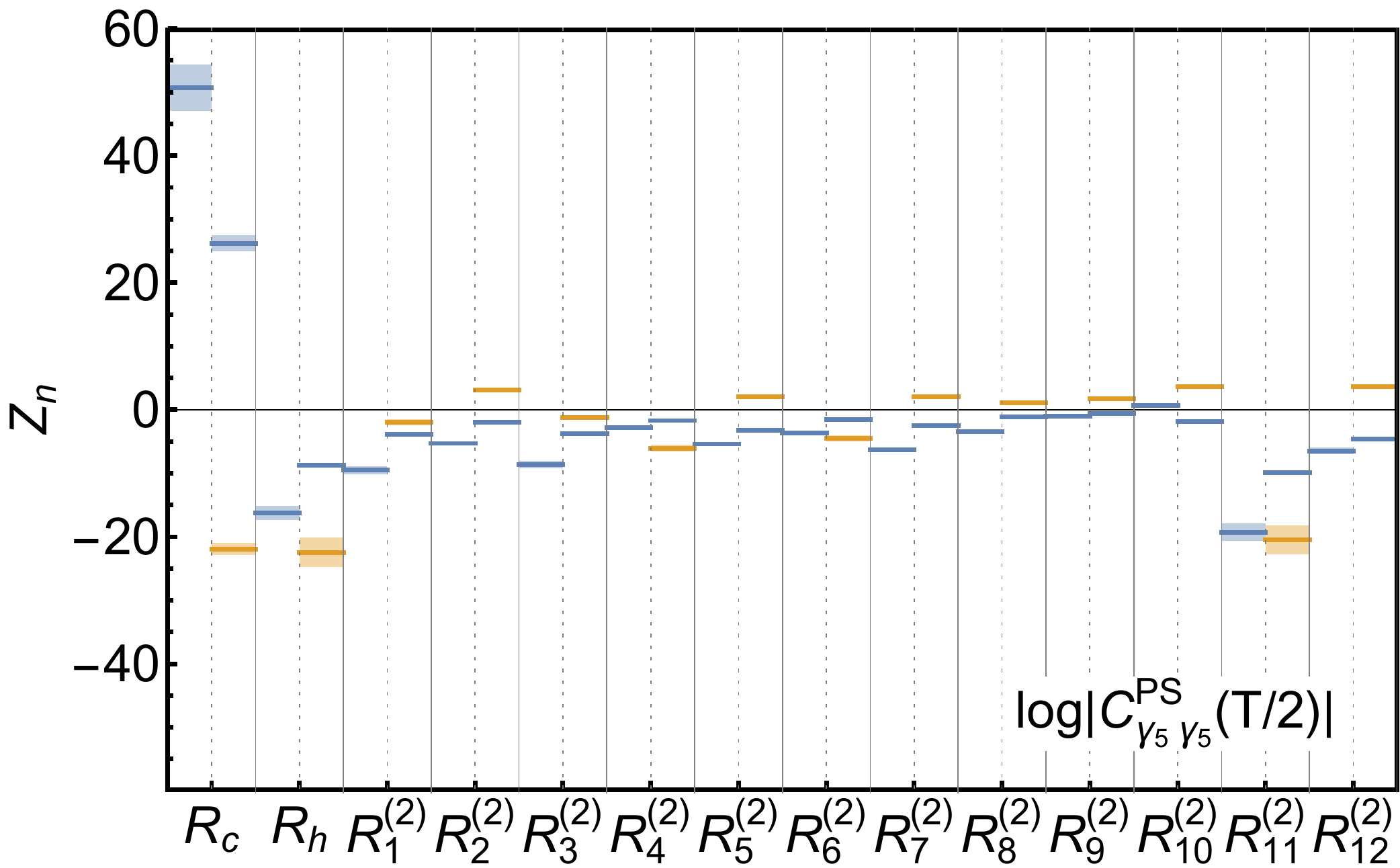}
\includegraphics[width=\figWidthHalf]{\figdir 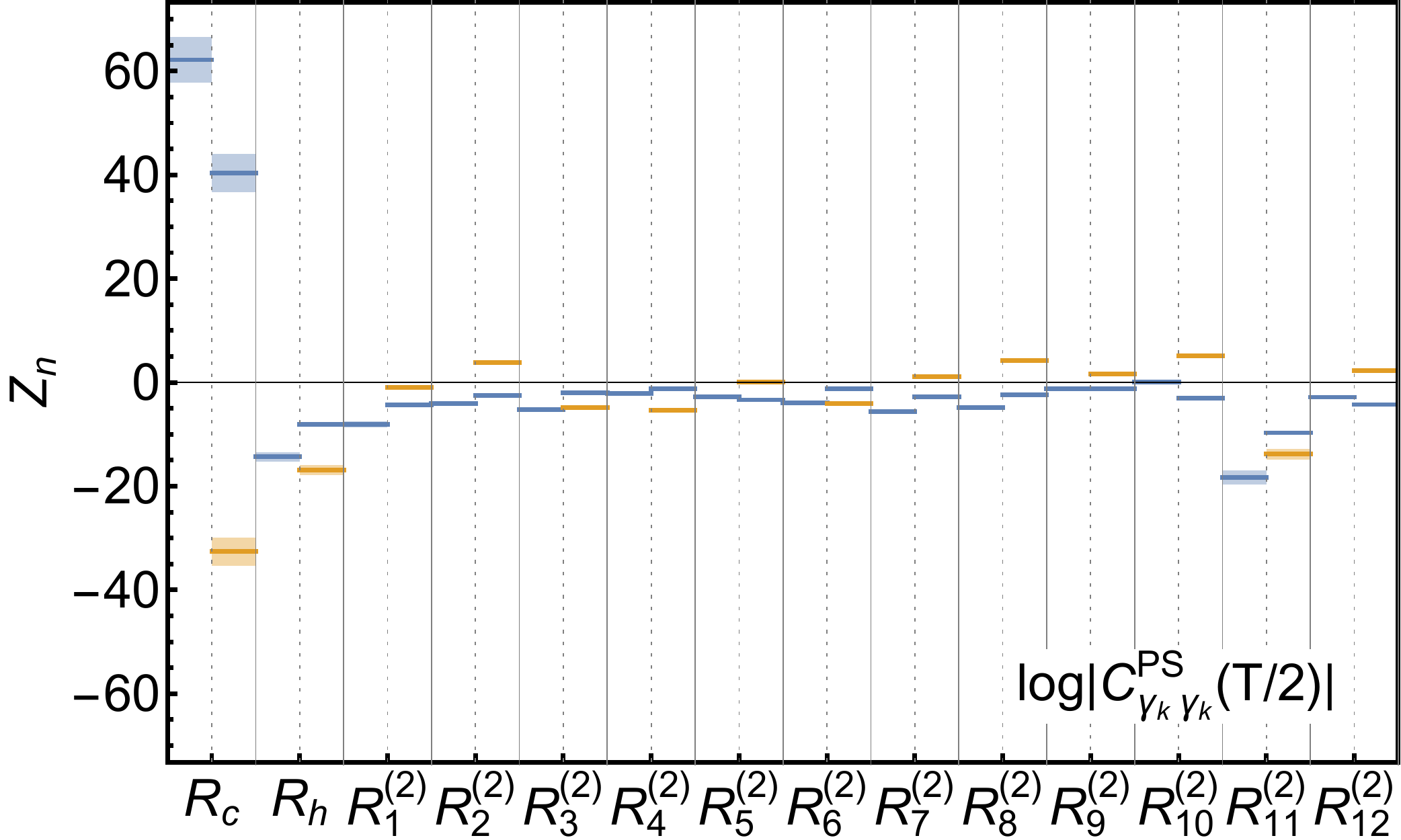}
\caption{\label{fig:therm_zFacts}% 
Overlap factors obtained from single- and double-exponential fits to thermalization curves for the plaquette (top, left), chiral condensate (top, right), action density at flow times $t_0/4$ (center, left) and $t_0$ (center, right), and pion (bottom, left) and rho (bottom, right) correlation functions evaluated at $T/2$.
Fit results correspond to the cases $n=1$ (blue) and $n=2$ (orange).
}
\end{figure}

\begin{figure}
\includegraphics[width=\figWidthHalf]{\figdir 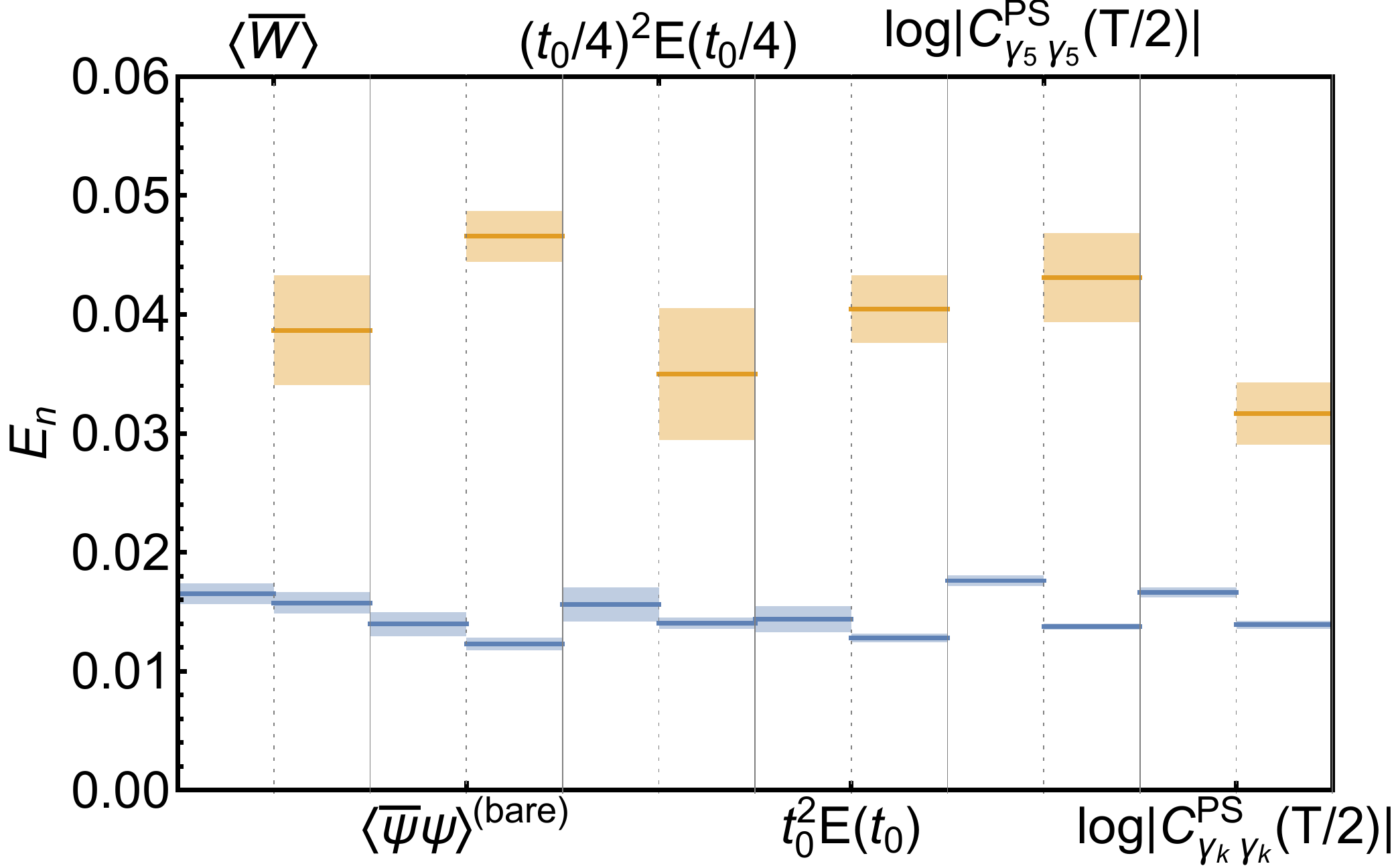} 
\caption{\label{fig:therm_en}% 
Thermalization time scales ($\tau_n = 1/E_n$) obtained from single- and double-exponential fits to thermalization curves for various observables.
Fit results correspond to the cases $n=1$ (blue) and $n=2$ (orange).
}
\end{figure}

Let $\calP_x$ ($x=c,h,1,\ldots,12$) represent the distributions from which the hot, cold, and prolongated coarse ensembles are drawn.
We extract the overlap factors $z_n(\calO,\calP_x)$ for various observables and initial probability distributions by performing coupled multiexponential fits to all data, using common exponents.
Single- and double-exponential fit results for $z_n(\calO,\calP_x)$ and $E_n$ are displayed in \Fig{therm_zFacts} and \Fig{therm_en}, respectively, for $n=1$ and $n=2$.
Although the uncertainties on each rethermalization stream are undetermined, we may use the fact that each stream is statistically independent in order to provide an estimate of the uncertainties in our fit parameters, assuming statistical fluctuations are independent of thermalization time.~\footnote{%
Since we are only using these rethermalization studies as a diagnostic of the efficacy of the RG matching procedure, we are not concerned with cleanly determining these algorithm dependent overlap parameters.
At sufficiently late times, the different starting configurations (hot, cold and prolongated) provide a thermalized fine ensemble of independent configurations and so the uncertainty on these physical quantities can be estimated appropriately, albeit with a small ensemble size.
Additionally, given that topological charge fluctuations are frequent for the fine couplings considered, one need not worry about systematic errors arising from topological freezing on a single stream.
} 
This is achieved by performing a bootstrap analysis over the independent streams, while keeping at least one sample corresponding to $F_h$ and $F_c$ within each bootstrap.
This constraint of the bootstrap procedure is imposed as many of the prolongated ensembles have only very short time windows in which any statistically significant Monte Carlo time dependence is apparent.
 Alternatively, one could use the hot and cold ensembles to determine the lowest two evolution eigenvalues and then use those fit results as inputs, fitting the overlaps of each of the prolongated ensembles onto these modes; similar conclusions would be reached.

\begin{figure}
\includegraphics[width=\figWidthHalf]{\figdir 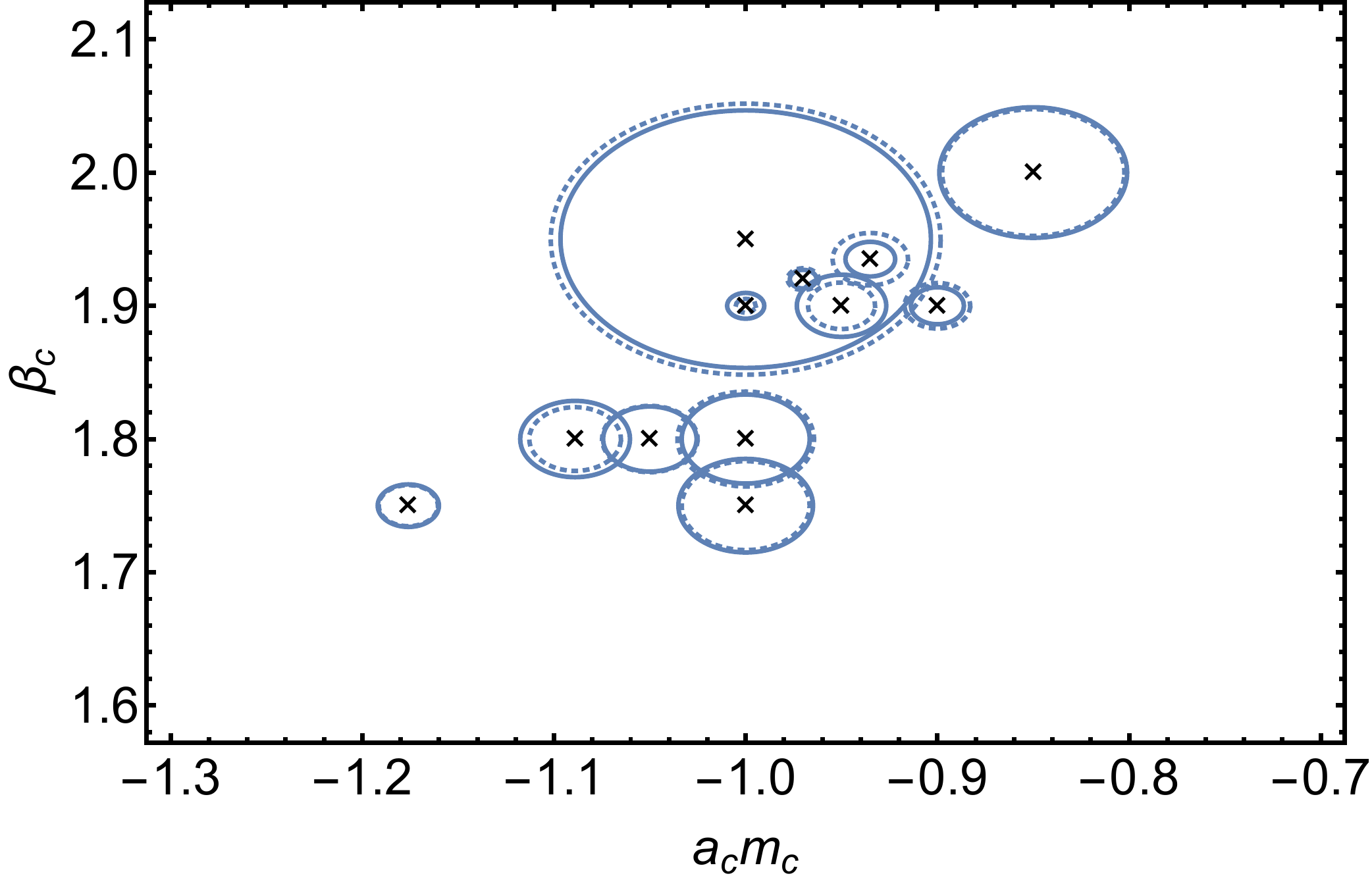} 
\caption{\label{fig:therm_z_summary}% 
Relationship between the coarse couplings used to produce the refined ensembles, $R^{(2)}_x$, and the resulting overlaps $|\langle \tilde \chi_1 | \calP_x \rangle|$ extracted from single-exponential (dotted) and double-exponential (solid) fit results for overlap factors $z_n$, displayed in \Fig{therm_zFacts}.
Circle radii illustrate the relative size of $|\langle \tilde \chi_1 | \calP_x \rangle|$ for various $x$, with smaller radii corresponding to smaller overlaps.
}
\end{figure}

In order to extract the relevant quantity, $\langle \tilde \chi_n | \calP_x \rangle$, from the overlap factors, we subsequently perform combined fits by minimizing
\begin{eqnarray}
\chi^2(a,b) = \sum_{\calO,x} \frac{[z_n(\calO,\calP_x) - a_\calO b_x]^2}{\sigma_{z_n}^2(\calO,\calP_x)}
\end{eqnarray}
for each $n$, subject to the normalization constraint $\sum_x |b_x| =1$.
Note that for a fixed $n$, the $\chi^2$ minimizing values of $a_\calO$ and $b_x$ provide best estimates for $\langle \calO | \chi_n \rangle$ and $\langle \tilde \chi_n | \calP_x \rangle$, respectively, up to an overall irrelevant normalization constant, which is fixed by the constraint.
A summary of the extracted overlaps $\langle \tilde \chi_1 | \calP_x \rangle$, and their association with each set of coarse ensemble parameters, are displayed in \Fig{therm_z_summary}.
From analysis of overlap factors, it is evident that $R_9^{(2)}$ and $R_{10}^{(2)}$ offer the fastest thermalization times, which is largely consistent with the RG analysis of \Sec{rg_matching}.
In principle, this information could be used to iteratively improve the RG matching before a large-scale set of calculations are undertaken. Importantly, the RG matching can be performed on ensembles of moderate volume, before undertaking full calculations on large volumes in which case significant effort can be invested in the tuning.

\section{Conclusion}

We have investigated a multiscale thermalization strategy for two-color QCD with two heavy fermion flavors.
We have demonstrated that a naive application of the methods proposed in~\cite{Endres:2015yca} leads to numerical instabilities in the rethermalization stage of HMC evolution; however, with minor modifications to the definition of the prolongator, these instabilities can be eliminated.
It is likely that there are other effective ways of eliminating the large fermion forces in the initial stages of rethermalization.
In particular, one might consider modifications of the time steps used in the evolution procedure (e.g., with the use of multiple time scale integration schemes~\cite{SEXTON1992665}), the use of higher-order integrators, or a modification of the fermion forces (e.g., with the use of mass~\cite{Hasenbusch:2001ne} and/or other preconditioning schemes).
Nevertheless, given the objective of providing initial configurations that rapidly thermalize under fine evolution, it is acceptable to use any prolongation procedure, such as the one that we adopt, providing it preserves the long-distance properties  of the configurations.

Although topological freezing is not an issue for the particular choice of couplings considered in this work, the multiscale method considered here provides an avenue for attaining ensembles with well-sampled topology in the limit of fine lattice spacing (albeit correctly distributed only up to inherited coarse action lattice artifacts).
It should be noted that a number of other promising methods have recently been proposed, which address the issue of topological freezing~\cite{Laio:2015era,Mages:2015scv}.
The method introduced in this work can be combined with those approaches to achieve further reductions in computational cost over those achieved by either individually.

The approach taken in this work is based on the premise that the configurations produced by prolongation have an underlying  distribution that is nearly orthogonal to the low modes of evolution.
Although there is no rigorous theoretical justification for this, intuitively, one expects this to be reasonable, providing the configurations originate from an ensemble generated using a well-matched coarse action, and the prolongated configurations inherit the long-distance features of the coarse configurations.
Previous studies of pure Yang-Mills theory indeed confirm this expectation.
In the case a two-color QCD with two heavy flavors of fermions, we find that matching both gluonic and fermionic observables result in rethermalization times that are significantly shorter than the typical thermalization times for ordered and disordered starts.
The results presented here are therefore an encouraging step forward in the application of multiscale methods to QCD simulations with three colors and physical quark masses.

\begin{acknowledgments}

We would like to thank Kostas Orginos for advice regarding numerical computations of the Dirac spectrum, and Richard Brower and Andrew Pochinsky for informative discussions.
All simulations were performed using a modified version of the Chroma Software System for lattice QCD~\cite{Edwards:2004sx}.
Determinations of the Dirac spectrum were performed using PRIMME (PReconditioned Iterative MultiMethod Eigensolver library)~\cite{Stathopoulos:2010:PPI:1731022.1731031}.
Computations for this study were carried out on facilities of the USQCD Collaboration, which are funded by the Office of Science of the U.~S.~Department of Energy. 
This work was partially supported by the U.~S.~Department of Energy through Early Career Research Award No. DE-SC0010495 and under Grant No. DE-SC0011090.
\end{acknowledgments}

\bibliography{qcd2}

%\appendix

\end{document}

%% file: tab1.tex
$C_{1}$ & 0.465135(32) & 0.311413(19) & 0.6064(2) & 0.31836(24)(15) & 0.2349(30)(18) & 1.2183(7)(2) & 1.3153(11)(9) \\
$C_{2}$ & 0.494151(76) & 0.315912(32) & 0.7761(10) & 0.06166(22)(19) & 0.1952(43)(73) & 0.6051(12)(19) & 0.8169(38)(20) \\
$C_{3}$ & 0.483510(34) & 0.309388(17) & 0.6541(3) & 0.26047(35)(4) & 0.2203(28)(19) & 1.1170(6)(6) & 1.2226(14)(11) \\
$C_{4}$ & 0.490134(40) & 0.311334(18) & 0.6894(4) & 0.18734(19)(6) & 0.2063(23)(18) & 0.9721(8)(12) & 1.1005(12)(13) \\
$C_{5}$ & 0.497182(75) & 0.312172(27) & 0.7352(8) & 0.12984(23)(22) & 0.2030(45)(29) & 0.8298(12)(7) & 0.9803(26)(48) \\
$C_{6}$ & 0.511431(40) & 0.300575(17) & 0.7415(4) & 0.27207(21)(40) & 0.1909(27)(24) & 1.1118(8)(4) & 1.1998(15)(6) \\
$C_{7}$ & 0.517712(53) & 0.302536(22) & 0.7917(6) & 0.19868(23)(42) & 0.1815(32)(27) & 0.9706(7)(5) & 1.0757(15)(18) \\
$C_{8}$ & 0.526442(54) & 0.303678(19) & 0.8855(11) & 0.12141(20)(22) & 0.1760(33)(23) & 0.7774(10)(10) & 0.9114(25)(23) \\
$C_{9}$ & 0.530302(77) & 0.301810(27) & 0.9003(15) & 0.13586(20)(14) & 0.1723(32)(19) & 0.8106(11)(9) & 0.9356(20)(14) \\
$C_{10}$ & 0.531018(66) & 0.300091(24) & 0.8814(12) & 0.16736(22)(4) & 0.1705(36)(23) & 0.8849(12)(6) & 0.9972(21)(26) \\
$C_{11}$ & 0.555057(157) & 0.298267(46) & 1.4471(139) & 0.02832(32)(39) & 0.1154(45)(61) & 0.4030(24)(13) & 0.6017(57)(62) \\
$C_{12}$ & 0.547748(61) & 0.293940(21) & 0.9947(20) & 0.19136(25)(35) & 0.1597(26)(50) & 0.9150(10)(29) & 1.0044(17)(16) \\
$F$ & 0.612219(15) & 0.280564(5) & 3.5075(140) & 0.06791(4)(14) & 0.1001(12)(13) & 0.4446(5)(3) & 0.5231(8)(18) \\

%% file: tab2.tex
$t_0/a^2$ & 10.38 & 115.10(68) & -94.04(46) & 59.89(57) & 19.65(8) & 8.79(10) & -23.48(20) \\
$a f_\pi$ & 0.00 & 5.51(2.72) & -3.47(1.63) & 3.60(2.79) & 0.55(27) & 0.69(54) & -1.14(96) \\
$a m_\pi$ & 0.53 & -91.49(1.17) & 73.01(73) & -59.26(1.09) & -14.57(12) & -10.56(21) & 22.40(38) \\
$a m_\rho$ & 0.34 & -72.06(2.02) & 58.46(1.32) & -46.66(1.75) & -11.83(23) & -8.31(37) & 17.77(56) \\

%% file: qcd2.bbl
\begin{thebibliography}{39}%
\makeatletter
\providecommand \@ifxundefined [1]{%
 \@ifx{#1\undefined}
}%
\providecommand \@ifnum [1]{%
 \ifnum #1\expandafter \@firstoftwo
 \else \expandafter \@secondoftwo
 \fi
}%
\providecommand \@ifx [1]{%
 \ifx #1\expandafter \@firstoftwo
 \else \expandafter \@secondoftwo
 \fi
}%
\providecommand \natexlab [1]{#1}%
\providecommand \enquote  [1]{``#1''}%
\providecommand \bibnamefont  [1]{#1}%
\providecommand \bibfnamefont [1]{#1}%
\providecommand \citenamefont [1]{#1}%
\providecommand \href@noop [0]{\@secondoftwo}%
\providecommand \href [0]{\begingroup \@sanitize@url \@href}%
\providecommand \@href[1]{\@@startlink{#1}\@@href}%
\providecommand \@@href[1]{\endgroup#1\@@endlink}%
\providecommand \@sanitize@url [0]{\catcode `\\12\catcode `\$12\catcode
  `\&12\catcode `\#12\catcode `\^12\catcode `\_12\catcode `\%12\relax}%
\providecommand \@@startlink[1]{}%
\providecommand \@@endlink[0]{}%
\providecommand \url  [0]{\begingroup\@sanitize@url \@url }%
\providecommand \@url [1]{\endgroup\@href {#1}{\urlprefix }}%
\providecommand \urlprefix  [0]{URL }%
\providecommand \Eprint [0]{\href }%
\providecommand \doibase [0]{http://dx.doi.org/}%
\providecommand \selectlanguage [0]{\@gobble}%
\providecommand \bibinfo  [0]{\@secondoftwo}%
\providecommand \bibfield  [0]{\@secondoftwo}%
\providecommand \translation [1]{[#1]}%
\providecommand \BibitemOpen [0]{}%
\providecommand \bibitemStop [0]{}%
\providecommand \bibitemNoStop [0]{.\EOS\space}%
\providecommand \EOS [0]{\spacefactor3000\relax}%
\providecommand \BibitemShut  [1]{\csname bibitem#1\endcsname}%
\let\auto@bib@innerbib\@empty
%</preamble>
\bibitem [{\citenamefont {Babich}\ \emph {et~al.}(2010)\citenamefont {Babich},
  \citenamefont {Brannick}, \citenamefont {Brower}, \citenamefont {Clark},
  \citenamefont {Manteuffel}, \citenamefont {McCormick}, \citenamefont
  {Osborn},\ and\ \citenamefont {Rebbi}}]{Babich:2010qb}%
  \BibitemOpen
  \bibfield  {author} {\bibinfo {author} {\bibfnamefont {R.}~\bibnamefont
  {Babich}}, \bibinfo {author} {\bibfnamefont {J.}~\bibnamefont {Brannick}},
  \bibinfo {author} {\bibfnamefont {R.~C.}\ \bibnamefont {Brower}}, \bibinfo
  {author} {\bibfnamefont {M.~A.}\ \bibnamefont {Clark}}, \bibinfo {author}
  {\bibfnamefont {T.~A.}\ \bibnamefont {Manteuffel}}, \bibinfo {author}
  {\bibfnamefont {S.~F.}\ \bibnamefont {McCormick}}, \bibinfo {author}
  {\bibfnamefont {J.~C.}\ \bibnamefont {Osborn}}, \ and\ \bibinfo {author}
  {\bibfnamefont {C.}~\bibnamefont {Rebbi}},\ }\href {\doibase
  10.1103/PhysRevLett.105.201602} {\bibfield  {journal} {\bibinfo  {journal}
  {Phys. Rev. Lett.}\ }\textbf {\bibinfo {volume} {105}},\ \bibinfo {pages}
  {201602} (\bibinfo {year} {2010})},\ \Eprint {http://arxiv.org/abs/1005.3043}
  {arXiv:1005.3043 [hep-lat]} \BibitemShut {NoStop}%
%%CITATION = ARXIV:1005.3043;%%
\bibitem [{\citenamefont {Babich}\ \emph {et~al.}(2009)\citenamefont {Babich},
  \citenamefont {Brannick}, \citenamefont {Brower}, \citenamefont {Clark},
  \citenamefont {Cohen}, \citenamefont {Osborn},\ and\ \citenamefont
  {Rebbi}}]{Babich:2009pc}%
  \BibitemOpen
  \bibfield  {author} {\bibinfo {author} {\bibfnamefont {R.}~\bibnamefont
  {Babich}}, \bibinfo {author} {\bibfnamefont {J.}~\bibnamefont {Brannick}},
  \bibinfo {author} {\bibfnamefont {R.~C.}\ \bibnamefont {Brower}}, \bibinfo
  {author} {\bibfnamefont {M.~A.}\ \bibnamefont {Clark}}, \bibinfo {author}
  {\bibfnamefont {S.~D.}\ \bibnamefont {Cohen}}, \bibinfo {author}
  {\bibfnamefont {J.~C.}\ \bibnamefont {Osborn}}, \ and\ \bibinfo {author}
  {\bibfnamefont {C.}~\bibnamefont {Rebbi}},\ }\bibfield  {booktitle} {\emph
  {\bibinfo {booktitle} {{Proceedings, 27th International Symposium on Lattice
  field theory (Lattice 2009)}}},\ }\href@noop {} {\bibfield  {journal}
  {\bibinfo  {journal} {PoS}\ }\textbf {\bibinfo {volume} {LAT2009}},\ \bibinfo
  {pages} {031} (\bibinfo {year} {2009})},\ \Eprint
  {http://arxiv.org/abs/0912.2186} {arXiv:0912.2186 [hep-lat]} \BibitemShut
  {NoStop}%
%%CITATION = ARXIV:0912.2186;%%
\bibitem [{\citenamefont {Frommer}\ \emph {et~al.}(2014)\citenamefont
  {Frommer}, \citenamefont {Kahl}, \citenamefont {Krieg}, \citenamefont
  {Leder},\ and\ \citenamefont {Rottmann}}]{Frommer:2013fsa}%
  \BibitemOpen
  \bibfield  {author} {\bibinfo {author} {\bibfnamefont {A.}~\bibnamefont
  {Frommer}}, \bibinfo {author} {\bibfnamefont {K.}~\bibnamefont {Kahl}},
  \bibinfo {author} {\bibfnamefont {S.}~\bibnamefont {Krieg}}, \bibinfo
  {author} {\bibfnamefont {B.}~\bibnamefont {Leder}}, \ and\ \bibinfo {author}
  {\bibfnamefont {M.}~\bibnamefont {Rottmann}},\ }\href {\doibase
  10.1137/130919507} {\bibfield  {journal} {\bibinfo  {journal} {SIAM J. Sci.
  Comput.}\ }\textbf {\bibinfo {volume} {36}},\ \bibinfo {pages} {A1581}
  (\bibinfo {year} {2014})},\ \Eprint {http://arxiv.org/abs/1303.1377}
  {arXiv:1303.1377 [hep-lat]} \BibitemShut {NoStop}%
%%CITATION = ARXIV:1303.1377;%%
\bibitem [{\citenamefont {Brannick}\ \emph {et~al.}(2015)\citenamefont
  {Brannick}, \citenamefont {Frommer}, \citenamefont {Kahl}, \citenamefont
  {Leder}, \citenamefont {Rottmann},\ and\ \citenamefont
  {Strebel}}]{Brannick:2014vda}%
  \BibitemOpen
  \bibfield  {author} {\bibinfo {author} {\bibfnamefont {J.}~\bibnamefont
  {Brannick}}, \bibinfo {author} {\bibfnamefont {A.}~\bibnamefont {Frommer}},
  \bibinfo {author} {\bibfnamefont {K.}~\bibnamefont {Kahl}}, \bibinfo {author}
  {\bibfnamefont {B.}~\bibnamefont {Leder}}, \bibinfo {author} {\bibfnamefont
  {M.}~\bibnamefont {Rottmann}}, \ and\ \bibinfo {author} {\bibfnamefont
  {A.}~\bibnamefont {Strebel}},\ }\href {\doibase 10.1007/s00211-015-0725-6}
  {\bibfield  {journal} {\bibinfo  {journal} {Numer. Math.}\ } (\bibinfo {year}
  {2015}),\ 10.1007/s00211-015-0725-6},\ \Eprint
  {http://arxiv.org/abs/1410.7170} {arXiv:1410.7170 [hep-lat]} \BibitemShut
  {NoStop}%
%%CITATION = ARXIV:1410.7170;%%
\bibitem [{\citenamefont {L{\"u}scher}\ and\ \citenamefont
  {Weisz}(2001)}]{Luscher:2001up}%
  \BibitemOpen
  \bibfield  {author} {\bibinfo {author} {\bibfnamefont {M.}~\bibnamefont
  {L{\"u}scher}}\ and\ \bibinfo {author} {\bibfnamefont {P.}~\bibnamefont
  {Weisz}},\ }\href {\doibase 10.1088/1126-6708/2001/09/010} {\bibfield
  {journal} {\bibinfo  {journal} {JHEP}\ }\textbf {\bibinfo {volume} {09}},\
  \bibinfo {pages} {010} (\bibinfo {year} {2001})},\ \Eprint
  {http://arxiv.org/abs/hep-lat/0108014} {arXiv:hep-lat/0108014 [hep-lat]}
  \BibitemShut {NoStop}%
%%CITATION = HEP-LAT/0108014;%%
\bibitem [{\citenamefont {Cè}\ \emph {et~al.}(2016)\citenamefont {Cè},
  \citenamefont {Giusti},\ and\ \citenamefont {Schaefer}}]{Ce:2016idq}%
  \BibitemOpen
  \bibfield  {author} {\bibinfo {author} {\bibfnamefont {M.}~\bibnamefont
  {Cè}}, \bibinfo {author} {\bibfnamefont {L.}~\bibnamefont {Giusti}}, \ and\
  \bibinfo {author} {\bibfnamefont {S.}~\bibnamefont {Schaefer}},\ }\href@noop
  {} {\  (\bibinfo {year} {2016})},\ \Eprint {http://arxiv.org/abs/1601.04587}
  {arXiv:1601.04587 [hep-lat]} \BibitemShut {NoStop}%
%%CITATION = ARXIV:1601.04587;%%
\bibitem [{\citenamefont {Vera}\ and\ \citenamefont
  {Schaefer}(2016)}]{Vera:2016xpp}%
  \BibitemOpen
  \bibfield  {author} {\bibinfo {author} {\bibfnamefont {M.~G.}\ \bibnamefont
  {Vera}}\ and\ \bibinfo {author} {\bibfnamefont {S.}~\bibnamefont
  {Schaefer}},\ }\href@noop {} {\  (\bibinfo {year} {2016})},\ \Eprint
  {http://arxiv.org/abs/1601.07155} {arXiv:1601.07155 [hep-lat]} \BibitemShut
  {NoStop}%
%%CITATION = ARXIV:1601.07155;%%
\bibitem [{\citenamefont {Goodman}\ and\ \citenamefont
  {Sokal}(1986)}]{Goodman:1986pv}%
  \BibitemOpen
  \bibfield  {author} {\bibinfo {author} {\bibfnamefont {J.}~\bibnamefont
  {Goodman}}\ and\ \bibinfo {author} {\bibfnamefont {A.~D.}\ \bibnamefont
  {Sokal}},\ }\href {\doibase 10.1103/PhysRevLett.56.1015} {\bibfield
  {journal} {\bibinfo  {journal} {Phys. Rev. Lett.}\ }\textbf {\bibinfo
  {volume} {56}},\ \bibinfo {pages} {1015} (\bibinfo {year}
  {1986})}\BibitemShut {NoStop}%
%%CITATION = PRLTA,56,1015;%%
\bibitem [{\citenamefont {Edwards}\ \emph {et~al.}(1991)\citenamefont
  {Edwards}, \citenamefont {Goodman},\ and\ \citenamefont
  {Sokal}}]{Edwards:1990hu}%
  \BibitemOpen
  \bibfield  {author} {\bibinfo {author} {\bibfnamefont {R.~G.}\ \bibnamefont
  {Edwards}}, \bibinfo {author} {\bibfnamefont {J.}~\bibnamefont {Goodman}}, \
  and\ \bibinfo {author} {\bibfnamefont {A.~D.}\ \bibnamefont {Sokal}},\ }\href
  {\doibase 10.1016/0550-3213(91)90357-4} {\bibfield  {journal} {\bibinfo
  {journal} {Nucl. Phys.}\ }\textbf {\bibinfo {volume} {B354}},\ \bibinfo
  {pages} {289} (\bibinfo {year} {1991})}\BibitemShut {NoStop}%
%%CITATION = NUPHA,B354,289;%%
\bibitem [{\citenamefont {Edwards}\ \emph {et~al.}(1992)\citenamefont
  {Edwards}, \citenamefont {Ferreira}, \citenamefont {Goodman},\ and\
  \citenamefont {Sokal}}]{Edwards:1991eg}%
  \BibitemOpen
  \bibfield  {author} {\bibinfo {author} {\bibfnamefont {R.~G.}\ \bibnamefont
  {Edwards}}, \bibinfo {author} {\bibfnamefont {S.~J.}\ \bibnamefont
  {Ferreira}}, \bibinfo {author} {\bibfnamefont {J.}~\bibnamefont {Goodman}}, \
  and\ \bibinfo {author} {\bibfnamefont {A.~D.}\ \bibnamefont {Sokal}},\ }\href
  {\doibase 10.1016/0550-3213(92)90262-A} {\bibfield  {journal} {\bibinfo
  {journal} {Nucl. Phys.}\ }\textbf {\bibinfo {volume} {B380}},\ \bibinfo
  {pages} {621} (\bibinfo {year} {1992})},\ \Eprint
  {http://arxiv.org/abs/hep-lat/9112002} {arXiv:hep-lat/9112002 [hep-lat]}
  \BibitemShut {NoStop}%
%%CITATION = HEP-LAT/9112002;%%
\bibitem [{\citenamefont {Janke}\ and\ \citenamefont
  {Sauer}(1994)}]{Janke:1993et}%
  \BibitemOpen
  \bibfield  {author} {\bibinfo {author} {\bibfnamefont {W.}~\bibnamefont
  {Janke}}\ and\ \bibinfo {author} {\bibfnamefont {T.}~\bibnamefont {Sauer}},\
  }\bibfield  {booktitle} {\emph {\bibinfo {booktitle} {{LATTICE 93: 11th
  International Symposium on Lattice Field Theory Dallas, Texas, October 12-16,
  1993}}},\ }\href {\doibase 10.1016/0920-5632(94)90509-6} {\bibfield
  {journal} {\bibinfo  {journal} {Nucl. Phys. Proc. Suppl.}\ }\textbf {\bibinfo
  {volume} {34}},\ \bibinfo {pages} {771} (\bibinfo {year} {1994})},\ \Eprint
  {http://arxiv.org/abs/hep-lat/9312043} {arXiv:hep-lat/9312043 [hep-lat]}
  \BibitemShut {NoStop}%
%%CITATION = HEP-LAT/9312043;%%
\bibitem [{\citenamefont {Grabenstein}\ and\ \citenamefont
  {Mikeska}(1994)}]{Grabenstein:1993nh}%
  \BibitemOpen
  \bibfield  {author} {\bibinfo {author} {\bibfnamefont {M.}~\bibnamefont
  {Grabenstein}}\ and\ \bibinfo {author} {\bibfnamefont {B.}~\bibnamefont
  {Mikeska}},\ }\bibfield  {booktitle} {\emph {\bibinfo {booktitle} {{LATTICE
  93: 11th International Symposium on Lattice Field Theory Dallas, Texas,
  October 12-16, 1993}}},\ }\href {\doibase 10.1016/0920-5632(94)90507-X}
  {\bibfield  {journal} {\bibinfo  {journal} {Nucl. Phys. Proc. Suppl.}\
  }\textbf {\bibinfo {volume} {34}},\ \bibinfo {pages} {765} (\bibinfo {year}
  {1994})},\ \Eprint {http://arxiv.org/abs/hep-lat/9311021}
  {arXiv:hep-lat/9311021 [hep-lat]} \BibitemShut {NoStop}%
%%CITATION = HEP-LAT/9311021;%%
\bibitem [{\citenamefont {Grabenstein}\ and\ \citenamefont
  {Pinn}(1994)}]{Grabenstein:1994ze}%
  \BibitemOpen
  \bibfield  {author} {\bibinfo {author} {\bibfnamefont {M.}~\bibnamefont
  {Grabenstein}}\ and\ \bibinfo {author} {\bibfnamefont {K.}~\bibnamefont
  {Pinn}},\ }\href {\doibase 10.1103/PhysRevD.50.6998} {\bibfield  {journal}
  {\bibinfo  {journal} {Phys. Rev.}\ }\textbf {\bibinfo {volume} {D50}},\
  \bibinfo {pages} {6998} (\bibinfo {year} {1994})},\ \Eprint
  {http://arxiv.org/abs/hep-lat/9406013} {arXiv:hep-lat/9406013 [hep-lat]}
  \BibitemShut {NoStop}%
%%CITATION = HEP-LAT/9406013;%%
\bibitem [{\citenamefont {Endres}\ \emph {et~al.}(2015)\citenamefont {Endres},
  \citenamefont {Brower}, \citenamefont {Detmold}, \citenamefont {Orginos},\
  and\ \citenamefont {Pochinsky}}]{Endres:2015yca}%
  \BibitemOpen
  \bibfield  {author} {\bibinfo {author} {\bibfnamefont {M.~G.}\ \bibnamefont
  {Endres}}, \bibinfo {author} {\bibfnamefont {R.~C.}\ \bibnamefont {Brower}},
  \bibinfo {author} {\bibfnamefont {W.}~\bibnamefont {Detmold}}, \bibinfo
  {author} {\bibfnamefont {K.}~\bibnamefont {Orginos}}, \ and\ \bibinfo
  {author} {\bibfnamefont {A.~V.}\ \bibnamefont {Pochinsky}},\ }\href {\doibase
  10.1103/PhysRevD.92.114516} {\bibfield  {journal} {\bibinfo  {journal} {Phys.
  Rev.}\ }\textbf {\bibinfo {volume} {D92}},\ \bibinfo {pages} {114516}
  (\bibinfo {year} {2015})},\ \Eprint {http://arxiv.org/abs/1510.04675}
  {arXiv:1510.04675 [hep-lat]} \BibitemShut {NoStop}%
%%CITATION = ARXIV:1510.04675;%%
\bibitem [{\citenamefont {Aoki}\ \emph {et~al.}(2006)\citenamefont {Aoki},
  \citenamefont {Blum}, \citenamefont {Christ}, \citenamefont {Dawson},
  \citenamefont {Izubuchi}, \citenamefont {Mawhinney}, \citenamefont {Noaki},
  \citenamefont {Ohta}, \citenamefont {Orginos}, \citenamefont {Soni},\ and\
  \citenamefont {Yamada}}]{PhysRevD.73.094507}%
  \BibitemOpen
  \bibfield  {author} {\bibinfo {author} {\bibfnamefont {Y.}~\bibnamefont
  {Aoki}}, \bibinfo {author} {\bibfnamefont {T.}~\bibnamefont {Blum}}, \bibinfo
  {author} {\bibfnamefont {N.~H.}\ \bibnamefont {Christ}}, \bibinfo {author}
  {\bibfnamefont {C.}~\bibnamefont {Dawson}}, \bibinfo {author} {\bibfnamefont
  {T.}~\bibnamefont {Izubuchi}}, \bibinfo {author} {\bibfnamefont {R.~D.}\
  \bibnamefont {Mawhinney}}, \bibinfo {author} {\bibfnamefont {J.}~\bibnamefont
  {Noaki}}, \bibinfo {author} {\bibfnamefont {S.}~\bibnamefont {Ohta}},
  \bibinfo {author} {\bibfnamefont {K.}~\bibnamefont {Orginos}}, \bibinfo
  {author} {\bibfnamefont {A.}~\bibnamefont {Soni}}, \ and\ \bibinfo {author}
  {\bibfnamefont {N.}~\bibnamefont {Yamada}},\ }\href {\doibase
  10.1103/PhysRevD.73.094507} {\bibfield  {journal} {\bibinfo  {journal} {Phys.
  Rev. D}\ }\textbf {\bibinfo {volume} {73}},\ \bibinfo {pages} {094507}
  (\bibinfo {year} {2006})}\BibitemShut {NoStop}%
\bibitem [{\citenamefont {Schaefer}\ \emph {et~al.}(2011)\citenamefont
  {Schaefer}, \citenamefont {Sommer},\ and\ \citenamefont
  {Virotta}}]{Schaefer:2010hu}%
  \BibitemOpen
  \bibfield  {author} {\bibinfo {author} {\bibfnamefont {S.}~\bibnamefont
  {Schaefer}}, \bibinfo {author} {\bibfnamefont {R.}~\bibnamefont {Sommer}}, \
  and\ \bibinfo {author} {\bibfnamefont {F.}~\bibnamefont {Virotta}} (\bibinfo
  {collaboration} {ALPHA}),\ }\href {\doibase 10.1016/j.nuclphysb.2010.11.020}
  {\bibfield  {journal} {\bibinfo  {journal} {Nucl. Phys.}\ }\textbf {\bibinfo
  {volume} {B845}},\ \bibinfo {pages} {93} (\bibinfo {year} {2011})},\ \Eprint
  {http://arxiv.org/abs/1009.5228} {arXiv:1009.5228 [hep-lat]} \BibitemShut
  {NoStop}%
%%CITATION = ARXIV:1009.5228;%%
\bibitem [{\citenamefont {Wilson}(1974)}]{PhysRevD.10.2445}%
  \BibitemOpen
  \bibfield  {author} {\bibinfo {author} {\bibfnamefont {K.~G.}\ \bibnamefont
  {Wilson}},\ }\href {\doibase 10.1103/PhysRevD.10.2445} {\bibfield  {journal}
  {\bibinfo  {journal} {Phys. Rev. D}\ }\textbf {\bibinfo {volume} {10}},\
  \bibinfo {pages} {2445} (\bibinfo {year} {1974})}\BibitemShut {NoStop}%
\bibitem [{\citenamefont {Duane}\ \emph {et~al.}(1987)\citenamefont {Duane},
  \citenamefont {Kennedy}, \citenamefont {Pendleton},\ and\ \citenamefont
  {Roweth}}]{DUANE1987216}%
  \BibitemOpen
  \bibfield  {author} {\bibinfo {author} {\bibfnamefont {S.}~\bibnamefont
  {Duane}}, \bibinfo {author} {\bibfnamefont {A.}~\bibnamefont {Kennedy}},
  \bibinfo {author} {\bibfnamefont {B.~J.}\ \bibnamefont {Pendleton}}, \ and\
  \bibinfo {author} {\bibfnamefont {D.}~\bibnamefont {Roweth}},\ }\href
  {\doibase http://dx.doi.org/10.1016/0370-2693(87)91197-X} {\bibfield
  {journal} {\bibinfo  {journal} {Physics Letters B}\ }\textbf {\bibinfo
  {volume} {195}},\ \bibinfo {pages} {216 } (\bibinfo {year}
  {1987})}\BibitemShut {NoStop}%
\bibitem [{\citenamefont {Banks}\ and\ \citenamefont
  {Casher}(1980)}]{Banks:1979yr}%
  \BibitemOpen
  \bibfield  {author} {\bibinfo {author} {\bibfnamefont {T.}~\bibnamefont
  {Banks}}\ and\ \bibinfo {author} {\bibfnamefont {A.}~\bibnamefont {Casher}},\
  }\href {\doibase 10.1016/0550-3213(80)90255-2} {\bibfield  {journal}
  {\bibinfo  {journal} {Nucl. Phys.}\ }\textbf {\bibinfo {volume} {B169}},\
  \bibinfo {pages} {103} (\bibinfo {year} {1980})}\BibitemShut {NoStop}%
%%CITATION = NUPHA,B169,103;%%
\bibitem [{\citenamefont {Edwards}\ \emph {et~al.}(1999)\citenamefont
  {Edwards}, \citenamefont {Heller},\ and\ \citenamefont
  {Narayanan}}]{Edwards:1998wx}%
  \BibitemOpen
  \bibfield  {author} {\bibinfo {author} {\bibfnamefont {R.~G.}\ \bibnamefont
  {Edwards}}, \bibinfo {author} {\bibfnamefont {U.~M.}\ \bibnamefont {Heller}},
  \ and\ \bibinfo {author} {\bibfnamefont {R.}~\bibnamefont {Narayanan}},\
  }\href {\doibase 10.1103/PhysRevD.59.094510} {\bibfield  {journal} {\bibinfo
  {journal} {Phys. Rev.}\ }\textbf {\bibinfo {volume} {D59}},\ \bibinfo {pages}
  {094510} (\bibinfo {year} {1999})},\ \Eprint
  {http://arxiv.org/abs/hep-lat/9811030} {arXiv:hep-lat/9811030 [hep-lat]}
  \BibitemShut {NoStop}%
%%CITATION = HEP-LAT/9811030;%%
\bibitem [{\citenamefont {Albanese}\ \emph
  {et~al.}(1987{\natexlab{a}})\citenamefont {Albanese} \emph
  {et~al.}}]{Albanese:1987ds}%
  \BibitemOpen
  \bibfield  {author} {\bibinfo {author} {\bibfnamefont {M.}~\bibnamefont
  {Albanese}} \emph {et~al.} (\bibinfo {collaboration} {APE}),\ }\href
  {\doibase 10.1016/0370-2693(87)91160-9} {\bibfield  {journal} {\bibinfo
  {journal} {Phys. Lett.}\ }\textbf {\bibinfo {volume} {B192}},\ \bibinfo
  {pages} {163} (\bibinfo {year} {1987}{\natexlab{a}})}\BibitemShut {NoStop}%
%%CITATION = PHLTA,B192,163;%%
\bibitem [{\citenamefont {Morningstar}\ and\ \citenamefont
  {Peardon}(2004)}]{Morningstar:2003gk}%
  \BibitemOpen
  \bibfield  {author} {\bibinfo {author} {\bibfnamefont {C.}~\bibnamefont
  {Morningstar}}\ and\ \bibinfo {author} {\bibfnamefont {M.~J.}\ \bibnamefont
  {Peardon}},\ }\href {\doibase 10.1103/PhysRevD.69.054501} {\bibfield
  {journal} {\bibinfo  {journal} {Phys. Rev.}\ }\textbf {\bibinfo {volume}
  {D69}},\ \bibinfo {pages} {054501} (\bibinfo {year} {2004})},\ \Eprint
  {http://arxiv.org/abs/hep-lat/0311018} {arXiv:hep-lat/0311018 [hep-lat]}
  \BibitemShut {NoStop}%
%%CITATION = HEP-LAT/0311018;%%
\bibitem [{\citenamefont {Groot}\ \emph {et~al.}(1984)\citenamefont {Groot},
  \citenamefont {Hoek},\ and\ \citenamefont {Smit}}]{Groot:1983ng}%
  \BibitemOpen
  \bibfield  {author} {\bibinfo {author} {\bibfnamefont {R.}~\bibnamefont
  {Groot}}, \bibinfo {author} {\bibfnamefont {J.}~\bibnamefont {Hoek}}, \ and\
  \bibinfo {author} {\bibfnamefont {J.}~\bibnamefont {Smit}},\ }\href {\doibase
  10.1016/0550-3213(84)90018-X} {\bibfield  {journal} {\bibinfo  {journal}
  {Nucl. Phys.}\ }\textbf {\bibinfo {volume} {B237}},\ \bibinfo {pages} {111}
  (\bibinfo {year} {1984})}\BibitemShut {NoStop}%
%%CITATION = NUPHA,B237,111;%%
\bibitem [{\citenamefont {Martinelli}\ and\ \citenamefont
  {Zhang}(1983)}]{Martinelli:1982mw}%
  \BibitemOpen
  \bibfield  {author} {\bibinfo {author} {\bibfnamefont {G.}~\bibnamefont
  {Martinelli}}\ and\ \bibinfo {author} {\bibfnamefont {Y.-C.}\ \bibnamefont
  {Zhang}},\ }\href {\doibase 10.1016/0370-2693(83)90987-5} {\bibfield
  {journal} {\bibinfo  {journal} {Phys. Lett.}\ }\textbf {\bibinfo {volume}
  {B123}},\ \bibinfo {pages} {433} (\bibinfo {year} {1983})}\BibitemShut
  {NoStop}%
%%CITATION = PHLTA,B123,433;%%
\bibitem [{\citenamefont {Meyer}\ and\ \citenamefont
  {Smith}(1983)}]{Meyer:1983ds}%
  \BibitemOpen
  \bibfield  {author} {\bibinfo {author} {\bibfnamefont {B.}~\bibnamefont
  {Meyer}}\ and\ \bibinfo {author} {\bibfnamefont {C.}~\bibnamefont {Smith}},\
  }\href {\doibase 10.1016/0370-2693(83)90959-0} {\bibfield  {journal}
  {\bibinfo  {journal} {Phys. Lett.}\ }\textbf {\bibinfo {volume} {B123}},\
  \bibinfo {pages} {62} (\bibinfo {year} {1983})}\BibitemShut {NoStop}%
%%CITATION = PHLTA,B123,62;%%
\bibitem [{\citenamefont {Detmold}\ \emph {et~al.}(2014)\citenamefont
  {Detmold}, \citenamefont {McCullough},\ and\ \citenamefont
  {Pochinsky}}]{Detmold:2014kba}%
  \BibitemOpen
  \bibfield  {author} {\bibinfo {author} {\bibfnamefont {W.}~\bibnamefont
  {Detmold}}, \bibinfo {author} {\bibfnamefont {M.}~\bibnamefont {McCullough}},
  \ and\ \bibinfo {author} {\bibfnamefont {A.}~\bibnamefont {Pochinsky}},\
  }\href {\doibase 10.1103/PhysRevD.90.114506} {\bibfield  {journal} {\bibinfo
  {journal} {Phys. Rev.}\ }\textbf {\bibinfo {volume} {D90}},\ \bibinfo {pages}
  {114506} (\bibinfo {year} {2014})},\ \Eprint {http://arxiv.org/abs/1406.4116}
  {arXiv:1406.4116 [hep-lat]} \BibitemShut {NoStop}%
%%CITATION = ARXIV:1406.4116;%%
\bibitem [{\citenamefont {L{\"u}scher}(2010)}]{Luscher:2010iy}%
  \BibitemOpen
  \bibfield  {author} {\bibinfo {author} {\bibfnamefont {M.}~\bibnamefont
  {L{\"u}scher}},\ }\href {\doibase 10.1007/JHEP08(2010)071,
  10.1007/JHEP03(2014)092} {\bibfield  {journal} {\bibinfo  {journal} {JHEP}\
  }\textbf {\bibinfo {volume} {1008}},\ \bibinfo {pages} {071} (\bibinfo {year}
  {2010})},\ \Eprint {http://arxiv.org/abs/1006.4518} {arXiv:1006.4518
  [hep-lat]} \BibitemShut {NoStop}%
%%CITATION = ARXIV:1006.4518;%%
\bibitem [{\citenamefont {L{\"u}scher}(1982)}]{Luscher:1981zq}%
  \BibitemOpen
  \bibfield  {author} {\bibinfo {author} {\bibfnamefont {M.}~\bibnamefont
  {L{\"u}scher}},\ }\href {\doibase 10.1007/BF02029132} {\bibfield  {journal}
  {\bibinfo  {journal} {Commun. Math. Phys.}\ }\textbf {\bibinfo {volume}
  {85}},\ \bibinfo {pages} {39} (\bibinfo {year} {1982})}\BibitemShut {NoStop}%
%%CITATION = CMPHA,85,39;%%
\bibitem [{\citenamefont {Phillips}\ and\ \citenamefont
  {Stone}(1986)}]{Phillips:1986qd}%
  \BibitemOpen
  \bibfield  {author} {\bibinfo {author} {\bibfnamefont {A.}~\bibnamefont
  {Phillips}}\ and\ \bibinfo {author} {\bibfnamefont {D.}~\bibnamefont
  {Stone}},\ }\href {\doibase 10.1007/BF01211167} {\bibfield  {journal}
  {\bibinfo  {journal} {Commun. Math. Phys.}\ }\textbf {\bibinfo {volume}
  {103}},\ \bibinfo {pages} {599} (\bibinfo {year} {1986})}\BibitemShut
  {NoStop}%
%%CITATION = CMPHA,103,599;%%
\bibitem [{\citenamefont {'t~Hooft}(1995)}]{'tHooft1995491}%
  \BibitemOpen
  \bibfield  {author} {\bibinfo {author} {\bibfnamefont {G.}~\bibnamefont
  {'t~Hooft}},\ }\href {\doibase
  http://dx.doi.org/10.1016/0370-2693(95)00251-F} {\bibfield  {journal}
  {\bibinfo  {journal} {Physics Letters B}\ }\textbf {\bibinfo {volume}
  {349}},\ \bibinfo {pages} {491 } (\bibinfo {year} {1995})}\BibitemShut
  {NoStop}%
\bibitem [{\citenamefont {Falcioni}\ \emph {et~al.}(1985)\citenamefont
  {Falcioni}, \citenamefont {Paciello}, \citenamefont {Parisi},\ and\
  \citenamefont {Taglienti}}]{Falcioni1985624}%
  \BibitemOpen
  \bibfield  {author} {\bibinfo {author} {\bibfnamefont {M.}~\bibnamefont
  {Falcioni}}, \bibinfo {author} {\bibfnamefont {M.}~\bibnamefont {Paciello}},
  \bibinfo {author} {\bibfnamefont {G.}~\bibnamefont {Parisi}}, \ and\ \bibinfo
  {author} {\bibfnamefont {B.}~\bibnamefont {Taglienti}},\ }\href {\doibase
  http://dx.doi.org/10.1016/0550-3213(85)90280-9} {\bibfield  {journal}
  {\bibinfo  {journal} {Nuclear Physics B}\ }\textbf {\bibinfo {volume}
  {251}},\ \bibinfo {pages} {624 } (\bibinfo {year} {1985})}\BibitemShut
  {NoStop}%
\bibitem [{\citenamefont {Albanese}\ \emph
  {et~al.}(1987{\natexlab{b}})\citenamefont {Albanese}, \citenamefont
  {Costantini}, \citenamefont {Fiorentini}, \citenamefont {Flore},
  \citenamefont {Lombardo}, \citenamefont {Tripiccione}, \citenamefont
  {Bacilieri}, \citenamefont {Fonti}, \citenamefont {Giacomelli}, \citenamefont
  {Remiddi}, \citenamefont {Bernaschi}, \citenamefont {Cabibbo}, \citenamefont
  {Marinari}, \citenamefont {Parisi}, \citenamefont {Salina}, \citenamefont
  {Cabasino}, \citenamefont {Marzano}, \citenamefont {Paolucci}, \citenamefont
  {Petrarca}, \citenamefont {Rapuano}, \citenamefont {Marchesini},\ and\
  \citenamefont {Rusack}}]{Albanese1987163}%
  \BibitemOpen
  \bibfield  {author} {\bibinfo {author} {\bibfnamefont {M.}~\bibnamefont
  {Albanese}}, \bibinfo {author} {\bibfnamefont {F.}~\bibnamefont
  {Costantini}}, \bibinfo {author} {\bibfnamefont {G.}~\bibnamefont
  {Fiorentini}}, \bibinfo {author} {\bibfnamefont {F.}~\bibnamefont {Flore}},
  \bibinfo {author} {\bibfnamefont {M.}~\bibnamefont {Lombardo}}, \bibinfo
  {author} {\bibfnamefont {R.}~\bibnamefont {Tripiccione}}, \bibinfo {author}
  {\bibfnamefont {P.}~\bibnamefont {Bacilieri}}, \bibinfo {author}
  {\bibfnamefont {L.}~\bibnamefont {Fonti}}, \bibinfo {author} {\bibfnamefont
  {P.}~\bibnamefont {Giacomelli}}, \bibinfo {author} {\bibfnamefont
  {E.}~\bibnamefont {Remiddi}}, \bibinfo {author} {\bibfnamefont
  {M.}~\bibnamefont {Bernaschi}}, \bibinfo {author} {\bibfnamefont
  {N.}~\bibnamefont {Cabibbo}}, \bibinfo {author} {\bibfnamefont
  {E.}~\bibnamefont {Marinari}}, \bibinfo {author} {\bibfnamefont
  {G.}~\bibnamefont {Parisi}}, \bibinfo {author} {\bibfnamefont
  {G.}~\bibnamefont {Salina}}, \bibinfo {author} {\bibfnamefont
  {S.}~\bibnamefont {Cabasino}}, \bibinfo {author} {\bibfnamefont
  {F.}~\bibnamefont {Marzano}}, \bibinfo {author} {\bibfnamefont
  {P.}~\bibnamefont {Paolucci}}, \bibinfo {author} {\bibfnamefont
  {S.}~\bibnamefont {Petrarca}}, \bibinfo {author} {\bibfnamefont
  {F.}~\bibnamefont {Rapuano}}, \bibinfo {author} {\bibfnamefont
  {P.}~\bibnamefont {Marchesini}}, \ and\ \bibinfo {author} {\bibfnamefont
  {R.}~\bibnamefont {Rusack}},\ }\href {\doibase
  http://dx.doi.org/10.1016/0370-2693(87)91160-9} {\bibfield  {journal}
  {\bibinfo  {journal} {Physics Letters B}\ }\textbf {\bibinfo {volume}
  {192}},\ \bibinfo {pages} {163 } (\bibinfo {year}
  {1987}{\natexlab{b}})}\BibitemShut {NoStop}%
\bibitem [{\citenamefont {de~Forcrand}\ \emph {et~al.}(1997)\citenamefont
  {de~Forcrand}, \citenamefont {Garcia~Perez},\ and\ \citenamefont
  {Stamatescu}}]{deForcrand:1997esx}%
  \BibitemOpen
  \bibfield  {author} {\bibinfo {author} {\bibfnamefont {P.}~\bibnamefont
  {de~Forcrand}}, \bibinfo {author} {\bibfnamefont {M.}~\bibnamefont
  {Garcia~Perez}}, \ and\ \bibinfo {author} {\bibfnamefont {I.-O.}\
  \bibnamefont {Stamatescu}},\ }\href {\doibase 10.1016/S0550-3213(97)00275-7}
  {\bibfield  {journal} {\bibinfo  {journal} {Nucl. Phys.}\ }\textbf {\bibinfo
  {volume} {B499}},\ \bibinfo {pages} {409} (\bibinfo {year} {1997})},\ \Eprint
  {http://arxiv.org/abs/hep-lat/9701012} {arXiv:hep-lat/9701012 [hep-lat]}
  \BibitemShut {NoStop}%
%%CITATION = HEP-LAT/9701012;%%
\bibitem [{\citenamefont {Sexton}\ and\ \citenamefont
  {Weingarten}(1992)}]{SEXTON1992665}%
  \BibitemOpen
  \bibfield  {author} {\bibinfo {author} {\bibfnamefont {J.}~\bibnamefont
  {Sexton}}\ and\ \bibinfo {author} {\bibfnamefont {D.}~\bibnamefont
  {Weingarten}},\ }\href {\doibase
  http://dx.doi.org/10.1016/0550-3213(92)90263-B} {\bibfield  {journal}
  {\bibinfo  {journal} {Nuclear Physics B}\ }\textbf {\bibinfo {volume}
  {380}},\ \bibinfo {pages} {665 } (\bibinfo {year} {1992})}\BibitemShut
  {NoStop}%
\bibitem [{\citenamefont {Hasenbusch}(2001)}]{Hasenbusch:2001ne}%
  \BibitemOpen
  \bibfield  {author} {\bibinfo {author} {\bibfnamefont {M.}~\bibnamefont
  {Hasenbusch}},\ }\href {\doibase 10.1016/S0370-2693(01)01102-9} {\bibfield
  {journal} {\bibinfo  {journal} {Phys. Lett.}\ }\textbf {\bibinfo {volume}
  {B519}},\ \bibinfo {pages} {177} (\bibinfo {year} {2001})},\ \Eprint
  {http://arxiv.org/abs/hep-lat/0107019} {arXiv:hep-lat/0107019 [hep-lat]}
  \BibitemShut {NoStop}%
%%CITATION = HEP-LAT/0107019;%%
\bibitem [{\citenamefont {Laio}\ \emph {et~al.}(2015)\citenamefont {Laio},
  \citenamefont {Martinelli},\ and\ \citenamefont {Sanfilippo}}]{Laio:2015era}%
  \BibitemOpen
  \bibfield  {author} {\bibinfo {author} {\bibfnamefont {A.}~\bibnamefont
  {Laio}}, \bibinfo {author} {\bibfnamefont {G.}~\bibnamefont {Martinelli}}, \
  and\ \bibinfo {author} {\bibfnamefont {F.}~\bibnamefont {Sanfilippo}},\
  }\href@noop {} {\  (\bibinfo {year} {2015})},\ \Eprint
  {http://arxiv.org/abs/1508.07270} {arXiv:1508.07270 [hep-lat]} \BibitemShut
  {NoStop}%
%%CITATION = ARXIV:1508.07270;%%
\bibitem [{\citenamefont {Mages}\ \emph {et~al.}(2015)\citenamefont {Mages},
  \citenamefont {Toth}, \citenamefont {Borsanyi}, \citenamefont {Fodor},
  \citenamefont {Katz},\ and\ \citenamefont {Szabo}}]{Mages:2015scv}%
  \BibitemOpen
  \bibfield  {author} {\bibinfo {author} {\bibfnamefont {S.}~\bibnamefont
  {Mages}}, \bibinfo {author} {\bibfnamefont {B.~C.}\ \bibnamefont {Toth}},
  \bibinfo {author} {\bibfnamefont {S.}~\bibnamefont {Borsanyi}}, \bibinfo
  {author} {\bibfnamefont {Z.}~\bibnamefont {Fodor}}, \bibinfo {author}
  {\bibfnamefont {S.}~\bibnamefont {Katz}}, \ and\ \bibinfo {author}
  {\bibfnamefont {K.~K.}\ \bibnamefont {Szabo}},\ }\href@noop {} {\  (\bibinfo
  {year} {2015})},\ \Eprint {http://arxiv.org/abs/1512.06804} {arXiv:1512.06804
  [hep-lat]} \BibitemShut {NoStop}%
%%CITATION = ARXIV:1512.06804;%%
\bibitem [{\citenamefont {Edwards}\ and\ \citenamefont
  {Joo}(2005)}]{Edwards:2004sx}%
  \BibitemOpen
  \bibfield  {author} {\bibinfo {author} {\bibfnamefont {R.~G.}\ \bibnamefont
  {Edwards}}\ and\ \bibinfo {author} {\bibfnamefont {B.}~\bibnamefont {Joo}}
  (\bibinfo {collaboration} {SciDAC, LHPC, UKQCD}),\ }\bibfield  {booktitle}
  {\emph {\bibinfo {booktitle} {{Lattice field theory. Proceedings, 22nd
  International Symposium, Lattice 2004, Batavia, USA, June 21-26, 2004}}},\
  }\href {\doibase 10.1016/j.nuclphysbps.2004.11.254} {\bibfield  {journal}
  {\bibinfo  {journal} {Nucl. Phys. Proc. Suppl.}\ }\textbf {\bibinfo {volume}
  {140}},\ \bibinfo {pages} {832} (\bibinfo {year} {2005})},\ \Eprint
  {http://arxiv.org/abs/hep-lat/0409003} {arXiv:hep-lat/0409003 [hep-lat]}
  \BibitemShut {NoStop}%
%%CITATION = HEP-LAT/0409003;%%
\bibitem [{\citenamefont {Stathopoulos}\ and\ \citenamefont
  {McCombs}(2010)}]{Stathopoulos:2010:PPI:1731022.1731031}%
  \BibitemOpen
  \bibfield  {author} {\bibinfo {author} {\bibfnamefont {A.}~\bibnamefont
  {Stathopoulos}}\ and\ \bibinfo {author} {\bibfnamefont {J.~R.}\ \bibnamefont
  {McCombs}},\ }\href {\doibase 10.1145/1731022.1731031} {\bibfield  {journal}
  {\bibinfo  {journal} {ACM Trans. Math. Softw.}\ }\textbf {\bibinfo {volume}
  {37}},\ \bibinfo {pages} {21:1} (\bibinfo {year} {2010})}\BibitemShut
  {NoStop}%
\end{thebibliography}%
